\documentclass[epsfig, usenatbib, usegraphicx]{mn2e}
\usepackage{epsfig}
\usepackage{natbib}
\usepackage{graphicx}
\title{Recovering galaxy star formation and metallicity histories from
  spectra using VESPA}
\author[Tojeiro et al.]{R. Tojeiro\thanks{E-mail: rmft@roe.ac.uk}$^1$,
  A. F. Heavens$^1$,  R. Jimenez$^2$ and B. Panter$^1$ \\
$^1$Institute for Astronomy, University of Edinburgh, Royal
Observatory, Blackford Hill, Edinburgh, EH9 3HJ, UK\\
$^2$Department of Physics and Astronomy, University of Pennsylvania, 
Philadelphia, PA-19104, USA
}

\def\gs{\mathrel{\raise1.16pt\hbox{$>$}\kern-7.0pt %
\lower3.06pt\hbox{{$\scriptstyle \sim$}}}}         %
\def\ls{\mathrel{\raise1.16pt\hbox{$<$}\kern-7.0pt %
\lower3.06pt\hbox{{$\scriptstyle \sim$}}}}         %

\newcommand{\aj}{AJ}

\newcommand{\mnras}{MNRAS}

\newcommand{\nat}{Nat}

\newcommand{\apj}{ApJ}
\newcommand{\apjs}{ApJ Supplement Series}
\newcommand{\pasp}{PASP}
\newcommand{\aaps}{A\&AS}
\begin{document}

\maketitle

\begin{abstract}
We introduce VErsatile SPectral Analysis (VESPA): a new method which
aims to
recover robust star formation and metallicity histories from galactic
spectra. VESPA uses the full spectral range to construct a galaxy history
from synthetic models. We investigate the use of an adaptative
parametrization grid to recover reliable star formation histories on
a galaxy-by-galaxy basis. Our goal is robustness as opposed to high
resolution histories, and the method is designed to return high time
resolution only where the data demand it. In this paper we detail the method and we
present our findings when we apply VESPA to synthetic and real Sloan
Digital Sky Survey (SDSS) spectroscopic data.  We show that the number of parameters that can be recovered from a
spectrum depends strongly on the signal-to-noise, wavelength coverage
and presence or absence of a young
population. For a typical SDSS sample of galaxies, we can normally  
recover between 2 to 5 stellar populations. 
We find very good agreement between VESPA and our previous analysis of the SDSS
sample with MOPED.
\end{abstract}


\title{Recovering galaxy histories using VESPA}

\section{Introduction}
\label{sec:introduction}
The spectrum of a galaxy holds vasts amounts of information about that
galaxy's history and evolution. Finding a way to tap directly into
this source of knowledge would not only provide us with crucial
information about that galaxy's evolutionary path, but would also
allow us to integrate this knowledge over a large number of galaxies
and therefore derive cosmological information. \\
\par\noindent
Galaxy formation and evolution are still far from being well
understood. Galaxies are extremely complex objects, formed via
complicated non-linear processes, and any approach (be it observational,
semi-analytical or computational) inevitably relies on
simplifications. If we try to analyse a galaxy's luminous output in
terms of a history parametrized by some chosen physical quantities,
such a 
simplification is also in order. The reason is two-fold: firstly
we are limited by our knowledge and ability to model all the physical
processes which happen in a galaxy and produce the observed spectrum
we are analysing; secondly, the observed spectrum is inevitably
perturbed by noise, which intrinsically limits the amount of
information we can recover.\\
\par\noindent
Measuring and understanding the star formation history of the Universe is therefore
essential to our understanding of galaxy evolution - when, where and
in what conditions did stars form throughout cosmic history? The
traditional and simplest way to probe this is to measure the observed
instantaneous star formation rate in galaxies at different
redshifts. This can be achieved by
looking at light emitted by young stars in the ultra-violet (UV)
band or its secondary effects. 
(e.g. \citealt{MadauEtAl96, Kennicutt98, HopkinsEtAl00, BundyEtAl06MNRAS, ErbEtAl06, AbrahamEtAl07MNRAS,
  NoeskeEtAl07MNRAS, VermaEtAl07}).
A complementary method is to look at present day galaxies and
extract their star formation history, which spans the lifetime of the
galaxy. Different teams have analysed a large number of galaxies in
this way, whether by using the full
available spectrum (\citealt{GlazebrookEtAl03MNRAS, PanterEtAl03, CidFernandesEtAl04, HeavensEtAl04,
MathisEtAl06, OcvirkEtAl06, PanterEtAl06, CidFernandesEtAl07}), or by
concentrating on particular spectral features or indices
(e.g. \citealt{KauffmannEtAl03aMNRAS, TremontiEtAl04MNRAS, GallazziEtAl05, BarberEtAl06}), which are
known to be correlated with age or metallicities
(e.g. \citealt{Worthey94, ThomasEtAl03}). \\
\par\noindent
To do this, we rely on synthetic stellar population
models to describe a galaxy in terms of its stellar components, 
but by modelling a galaxy in
this way we are intrinsically limited by the quality of the models.
There are also potential concerns with flux calibration
errors. However, using the full spectrum to
recover the fossil record of a galaxy - or of an ensemble of galaxies
- is an
extremely powerful method, as the quality and amount of data relating to local
galaxies vastly outshines that which concerns high-redshift galaxies. Splitting a
galaxy into simple stellar populations of different ages and
metallicities is a natural way of parameterising a galaxy, and it
allows realistic fits to real galaxies
(e.g. \citealt{BruzualEtCharlot03}). Galactic
archeology has become increasingly popular in the literature recently,
largely due to the increase in
sophistication of stellar population synthesis codes and the
improvement of the stellar spectra libraries upon which they are
based, and also due to the availability of large well-calibrated
spectroscopic databases, such as the Sloan Digital Sky Survey (SDSS)
(\citealt{YorkEtAl00MNRAS, StraussEtAl02MNRAS}).\\
\par\noindent
In any case, without imposing any
constraints on the allowed form of the star formation history, or perhaps an age-metallicity
relation, the parameter space can become unsustainably large for a
traditional approach. Ideally, one would like to do without such
pre-constraints. Recently, different research teams have come up with
widely different solutions for this problem. MOPED \citep{HeavensJimenezLahav00} and
STARLIGHT \citep{CidFernandesEtAl04} explore a well-chosen parameter space in order to find
the best possible fit to the data. In the case of
MOPED, this relies on compression of the full spectrum to a much smaller
set of numbers which retains all the information about the
parameters it tries to recover; STARLIGHT on the other hand, searches
for its best fit using the full spectrum with a Metropolis algorithm. STECMAP \citep{OcvirkEtAl06}
solves the problem using an algebraic least-squares solution and a
well-chosen regularization to keep the inversions stable. All of
these and other methods
acknowledge the same limitation - noise in the data and in the models
introduces degeneracies into the
problem which can lead to unphysical results. MOPED, for example, has
produced some remarkable results concerning the average star formation
history of the Universe by analysing a large sample of
galaxies. However, MOPED's authors have cautioned against
over-interpreting the results on a galaxy-by-galaxy basis, due to the problem mentioned
above. This is directly related to the question of how finely one should
parameterise a galaxy, and what the consequences of this might be.\\
\par\noindent
Much of the motivation for VESPA came
from the realisation that this problem will vary from galaxy to
galaxy, and that the method of choosing a single parametrization to
analyse a large
number of galaxies can be improved on. \\
\par\noindent
VESPA is based on three main ideas, which we present here and develop
further in the main text: 
\begin{itemize}
\item There is only so much information one can safely recover from
  any given set of data, and the amount of information which can  be
  recovered from an individual galaxy varies.
\item The recovered star formation fractions should be positive. 
\item Even though the full unconstrained problem is non-linear, it is
  piecewise linear in well-chosen regions of parameter space.
\end{itemize}
\par\noindent
VESPA's ultimate goal is to derive robust information for
each galaxy individually, by adapting the number of parameters it
recovers on a galaxy-by-galaxy basis and increasing the resolution in
parameter space only where the data warrant it. In a nutshell, this
is how VESPA works: we estimate how many parameters we can recover from
a given spectrum, given its noise, shape, spectral resolution and
wavelength range using an analysis given by \cite{OcvirkEtAl06}. In that paper,
Singular Value Decomposition (SVD) is used to find a least squares solution,
and this solution is analysed in terms of its singular vectors. VESPA uses
this method only as an analysis of the solution, and uses Bounded-Variable Least-Squares (BVLS) \citep{StarkAndParker95} to reach a
non-negative solution in several regimes where linearity applies.  \\
\par\noindent
This paper is organised as follows: in Section \ref{sec:method} we
present the method, in Section \ref{sec:tests} we apply VESPA to
a variety of synthetic spectra, in Section \ref{sec:results} we apply
VESPA to a sample of galaxies from the Sloan Digital Sky Survey
spectroscopic database and we compare our results to those obtained
with MOPED, and finally in Section \ref{sec:conclusions} we present our
conclusions. 


\section{Method}
\label{sec:method}
In this section we lay down
the problem to solve in detail, and explain the different steps VESPA
uses to find a solution for each galaxy.

\subsection{The problem}

We assume a galaxy is composed of a series of simple stellar
populations (SSP) of varying ages and metallicities. The unobscured rest frame
luminosity per unit wavelength of a galaxy can then be written as
\begin{equation}
F_{\lambda} = \int^t_0 \psi(t) S_{\lambda}(t, Z)dt
\label{eq:prob_int}
\end{equation}
\par\noindent
where $\psi(t)$ is the star formation rate (solar masses formed per
unit of time) and $S_{\lambda}(t,Z)$ is the luminosity per unit
wavelength of a single stellar
population of age $t$ and metallicity $Z$, per unit mass. The
dependency of the metallicity on age is unconstrained, turning this into a
non-linear problem.\\
\par\noindent
In order to solve this problem, we start by discretizing in wavelength and time,
by averaging these two quantities into well chosen bins. For now we
present the problem with a generalised parametrization, and discuss
our choice in Section \ref{sec:parametrization}. We will use greek
indices to indicate time bins, and roman indices to indicate
wavelength bins. \\
\par\noindent
The problem becomes

\begin{equation}
F_j = \sum_\alpha x_\alpha G(Z_\alpha)_{\alpha j} 
\end{equation}
\par\noindent
where $F_j=(F_1, ..., F_D)$ is the luminosity of the $j$th wavelength
bin of width $\Delta \lambda$, $G(Z_\alpha)_{\alpha j}$ is
the $j$th luminosity point of a stellar population of age $t_\alpha =(t_1,...,t_S)$
(spanning an age range of $\Delta t_\alpha$) and metallicity $Z_\alpha$, and
$x_\alpha = (x_1, ..., x_S)$ is the total mass of population
$G(Z)_{\alpha j}$ in the
time bin $\Delta t_\alpha$.\\
\par\noindent
Although the full metallicity problem is non-linear, interpolating
between tabulated values of $Z$ gives a piecewise linear behaviour:
\begin{equation}
G(Z_\alpha)_{\alpha j} = g_\alpha G(Z_{a,\alpha})_{\alpha j} +
(1-g_\alpha)G(Z_{b,\alpha})_{\alpha j},
\end{equation}
and the problem then becomes
\begin{equation}
F_j = \sum_\alpha x_\alpha \left[ g_\alpha G(Z_{a,\alpha})_{\alpha j} + \left(1 - g_\alpha\right)
  G(Z_{b,\alpha})_{\alpha j} \right]
\label{eq:linear}
\end{equation}
\par\noindent
where $G(Z_{a,\alpha})_{\alpha j}$ and $G(Z_{b,\alpha})_{\alpha j}$
are equivalent to $G(Z_\alpha)_{\alpha j}$ as
above, but at fixed metallicities $Z_{a}$ and $Z_{b}$, which bound the
true Z. If this interpolation matches the models' resolution in Z,
then we are not degrading the models in any way.\\
\par\noindent
Solving the problem then requires finding the correct metallicity
range. One should not underestimate the complexity this implies - 
trying all possible combination of consecutive values of $Z_a$
and $Z_b$ in a grid of 16 age bins would
lead to a total number of calculations of the order of $10^9$, which
is unfeasible even in today's fast personal workstations. We work
around this problem using an iterative approach, which we describe in 
Section \ref{sec:search}. 

\subsubsection{Dust extinction}
\label{sec:dust}
An important component when describing the luminous output of a
galaxy is dust, as different wavelengths are affected in different
ways. The simplest possible approach is to use one-parameter dust
model, according to which we apply a single dust screen to the
combined luminosity of all the galactic components. Equation (\ref{eq:prob_int}) becomes
\begin{equation}
F_{\lambda} = f_{dust}(\tau_\lambda)\int^t_0 \psi(t) S_{\lambda}(t, Z)dt
\end{equation}
\par\noindent
where we are assuming the dust extinction is the same for all stars,
and characterised by the optical depth, $\tau_\lambda$. \\
\par\noindent
However, it is also well known that very young stars are likely to be
more affected by dust. In an attempt to include this in our modelling,
we follow the two-parameter dust model of \cite{CharlotFall00} in
which young stars are embebbed in their birth cloud up to a time
$t_{BC}$, when they break free into the inter-stellar medium (ISM):

\begin{equation}
F_{\lambda} = \int_0^t f_{dust}(\tau_\lambda, t) \psi(t) S_{\lambda}(t, Z)dt
\end{equation}
and
\begin{equation}
f_{dust}(\tau_\lambda, t) = \left\{
\begin{array}{l}
f_{dust}(\tau^{ISM}_\lambda) f_{dust}(\tau^{BC}_\lambda), t \leq t_{BC}\\
f_{dust}(\tau^{ISM}_\lambda), t > t_{BC}\\
\end{array}
\right.
\end{equation}
where $\tau_\lambda^{ISM}$ is the optical depth of the ISM and
$\tau_\lambda^{BC}$ is the optical depth of the birth cloud. Following
\cite{CharlotFall00}, we take $t_{BC}= 0.03$ Gyrs.\\
\par\noindent
There is a variety of choices for the form of
$f_{dust}(\tau_\lambda)$. To model the dust in the ISM, we use the mixed slab model of
\cite{CharlotFall00} for low optical depths ($\tau_V \le 1$), for which
\begin{equation}
f_{dust}(\tau_\lambda) = \frac{1}{2\tau_\lambda}[1 + (\tau_\lambda -
1)\exp(-\tau_\lambda) - \tau_\lambda^2E_1(\tau_\lambda)]
\end{equation}
where $E_1$ is the exponential integral and
$\tau_\lambda$ is the optical depth of the slab. This model is known
to be less accurate for high dust values, and for optical depths
greater than one we take a uniform screening model with
\begin{equation}
f_{dust}(\tau_\lambda) = \exp(-\tau_\lambda).
\end{equation}
We only use the uniform screening model to model the dust in the birth
cloud and we use $\tau_\lambda =\tau_V(\lambda/5500\AA)^{-0.7}$ as our
extinction curve for both environments.\\
\par\noindent
As described, dust is a non-linear problem. In practice, we solve the linear
problem described by equation (\ref{eq:linear}) with a number of dust
extinctions applied to the matrices $G(Z)_{ij}$ and choose the
values of $\tau_V^{ISM}$ and $\tau_V^{BC}$ which result in the best
fit to the data. \\
\par\noindent
We initially use a
binary chop search for $\tau_V^{ISM} \in [0,4]$ and keep $\tau_V^{BC}$
fixed and equal to zero, which results in trying out
typically around nine values of $\tau_V^{ISM}$. If this initial
solution reveals star formation at a time less than $t_{BC}$ we repeat
our search on a two-dimensional grid, and fit for $\tau_V^{ISM}$ and
$\tau_V^{BC}$ simultaneously. There is no penalty except in CPU time to apply the two-parameter search, but we find that this procedure is robust (see section 3.4).

\subsection{The solution}
\label{sec:solution}
In this section we describe the method used to reach a solution for a
galaxy, given a set of models and a generalised parametrization. The
construction of these models and choice of parameters is
described in Sections \ref{sec:parametrization} and \ref{subsec:models}.\\
\par\noindent
We re-write the problem described by equation (\ref{eq:linear}) in a
simpler way
\begin{equation}
F_j = \sum_{\kappa=1}^{2S} c_\kappa A_{\kappa j}(Z_\kappa)
\label{eq:matrix}
\end{equation}
\par\noindent
where $Z_\kappa = Z_a$ for $\kappa<S$ and $Z_\kappa=Z_b$ for $\kappa \ge S$.
$\bf{A}$ is a $D \times 2S$ matrix composed of synthetic models at the
corresponding metallicities, and ${\bf c}=(c_1, ..., c_{2S})$ is the
solution vector, from which the $x_\alpha$ and $g_\alpha$ in equation
(\ref{eq:linear}) can be calculated. We can then calculate a linearly
interpolated metallicity at age $t_\alpha$

\begin{equation}
Z_\alpha = g_\alpha Z_a + (1 - g_\alpha) Z_b.
\end{equation}
\par\noindent
For every age $t_\alpha$ we aim to recover two parameters: $x_\alpha$ - the total
mass formed at that age (within a period $\Delta t_\alpha$) - and $Z_\alpha$ - a
mass-weighted metallicity.\\
\par\noindent
At this stage we are not concerned with our choice of
$t_\alpha$ and $\Delta t_\alpha$ - although these are crucial and will be
discussed later. 
For a given set of chosen parameters, we find ${\bf c}$, such that

\begin{equation}
\chi^2 = \frac{( F_j - \sum_{\kappa}c_\kappa A_{\kappa j})^2}{\sigma_j^2}
\end{equation}
\par\noindent
is minimised (where $\sigma_j$ is the error in the measured flux bin $F_j$). \\
\par\noindent
A linear problem with a least squares constraint has a simple analytic solution
which, for constant $\bf{\sigma_j}$ (white-noise) is
\begin{equation}
\bf{c}_{LS} = (\bf{A}^T \cdot \bf{A})^{-1}\bf{\cdot}\bf{A}^T\bf{\cdot F} 
\label{eq:inversion}
\end{equation}
\par\noindent
In principle, any matrix inversion method, e.g. Singular Value
Decomposition (SVD), can be used to solve
(\ref{eq:inversion}). However, we would like to impose positivity
constraints on the recovered solutions. Negative solutions are
unphysical, but unfortunately common in a problem perturbed by
noise. 

\subsubsection{BVLS and positivity}
\label{sec:bvls}
We use Bounded-Variable Least-Squares (BVLS)
\citep{StarkAndParker95} in order to solve
(\ref{eq:inversion}). 
BVLS is an algorithm which solves linear problems whose
variables are subject to upper and lower constraints. It uses an active set
strategy to choose sets of free variables, and uses QR
matrix decomposition to solve the unconstrained least-squares problem
of each candidate set of free variables using (\ref{eq:inversion}):
\begin{equation}
\bf{c_{LS}} = (\bf{E}^T \cdot \bf{E})^{-1}\bf{\cdot E}^T\bf{\cdot F}
\end{equation}
where $\bf{E}$ is effectively composed of those columns of $\bf{A}$
for which $c_k$ is unconstrained, and of zero vectors for those columns
for which $c_k$ is set to zero. BVLS is an extension of the
Non-Negative Least Squares algorithm \citep{LawsonBook}, and they are both proven to
converge in a finite number of iterations. 
Positivity is the only constraint in VESPA's solution.\\
\par\noindent
BVLS and positivity have various advantages. Most obvious is the fact
that we do away with negative solutions. In a non-constrained method
(such as SVD) negative values are a response to the fact that the data
is noisy. Similarly, we find that zero values returned by BVLS (in,
for example, a synthetic galaxy with continuous star formation across
all time) are also an artifact from noisy data. It should be kept in
mind that, if the method is unbiased, this problem is solved by
analysing a number of noisy realisations of the original problem -
what we find is that the true values of the parameters we try to
recover are consistent with the distributions yielded by this
process. In this sense, not even a negative value presents a problem
necessarily, as long as it is consistent with zero (or the correct
solution). Given that we have found no bias when using BVLS, we feel
it is an advantage to discard a priori solutions we know to be
unphysical. \\
\par\noindent
Another advantage to using BVLS is the fact that, by fixing some
parameters to the lower boundary (zero, in this case), it effectively
reduces the number of fitting parameters to the number of those which keeps
unconstrained. Given the overall aims of VESPA, this has proven to be
advantageous. 

\subsubsection{Noise}
\label{sec:noise_th}

The inversion in equation (\ref{eq:inversion}) is often highly
sensitive to noise, and care is needed when recovering solutions with 
matrix inversion methods. The fit in data-space will always improve as
we increase the number of parameters, but these might not all provide
meaningful information. We follow an analysis given in \cite{OcvirkEtAl06} in order to
understand how much this affects our results, and to choose a suitable
age parametrization for each galaxy. This is not an exact method, and
it does not guarantee that the solutions we recover have no
contribution from noise. However, we found that in most cases it
provides a very useful guideline (see section \ref{sec:noise}, in
particular Figure \ref{fig:svdtest}). \\
\par\noindent
We refer the reader to the above
paper for a full discussion, and we reproduce here the steps used in
our analysis. \\
\par\noindent
We use SVD to decompose the model matrix $\bf{E}$ as
\begin{equation}
\bf{E} = \bf{U \cdot W \cdot V}^T
\end{equation}
\par\noindent
where $\bf{U}$ is a $D \times 2S$ orthonormal matrix with
singular data-vectors $\bf{u}_\kappa$ as columns,
$\bf{V}$ is a $2S \times 2S$ orthonormal matrix with the singular
solution-vectors $\bf{v}_\kappa$ as columns, and ${\bf W}$ is a $2S \times 2S$ diagonal
matrix ${\bf W} = diag(w_1, ..., w_{2S})$ where $w_\kappa$ are the matrix
singular values in decreasing order. Replacing $\bf{E}$ by this
decomposition in equation (\ref{eq:inversion}) gives
\begin{equation}
{\bf c}_{LS} = {\bf V \cdot W}^{-1}{\bf \cdot U}^T{\bf \cdot F} = \sum_{\kappa=1}^{2S}
\frac{{\bf u}_\kappa^T {\bf \cdot F}}{w_\kappa}{\bf v}_\kappa
\label{eq:svd}
\end{equation}
\par\noindent
The solution vector is a linear combination of the solution singular
values, parametrized by the dot product between the data and the
corresponding data singular vector, and divided by the $k^{th}$
singular value. The data vector itself is a combination of the true underlying emitted flux
and noise: $\bf{F = F_{true} + e}$. Equation (\ref{eq:svd}) becomes
\begin{equation}
{\bf c}_{LS} = \sum_{\kappa=1}^{2S} \frac{{\bf u}_\kappa^T {\bf \cdot F}_{true}}{
      w_\kappa}{\bf v}_\kappa
+ \sum_{\kappa=1}^{2S} \frac{{\bf u}_\kappa^T {\bf \cdot e}}{ w_\kappa}{\bf v}_\kappa \equiv
{\bf c}_{true} + {\bf c}_e
\label{eq:svd_sum}
\end{equation}
\par\noindent
where $\bf{c}_{true}$ is the solution vector to the noiseless problem
and $\bf{c}_e$ is an unavoidable added term due to the presence of
noise.\\

\begin{figure}
\vspace{0.6cm}
\begin{center}
\includegraphics[width=3.in]{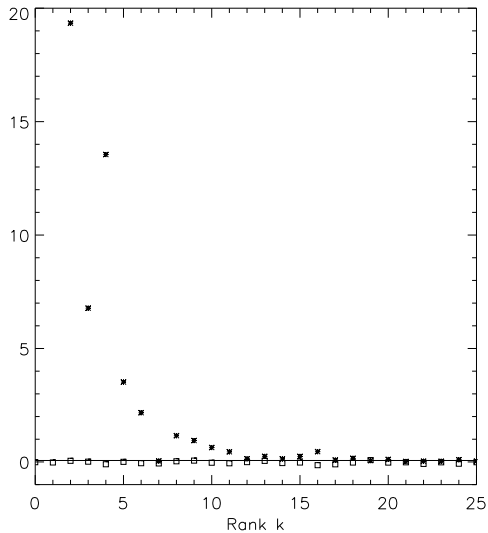}
\caption{The behaviour of the singular values with matrix rank
  $k$. The stars are $|{\bf u}_\kappa^T {\bf \cdot F}|$ and the squares are ${\bf
  u}_\kappa^T {\bf \cdot e}$. The line is $\left\langle{\bf
  F}\right\rangle/$SNR, which in this case has a value of approximately 0.06. \small }

\label{fig:sing_values}
\end{center}
\end{figure}

\par
\begin{figure}
\vspace{0.6cm}
\begin{center}
\includegraphics[width=3.in]{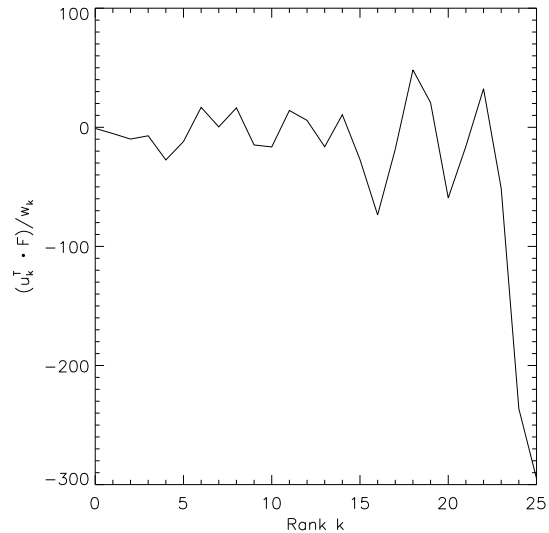}
\caption{The coefficients in sum (\ref{eq:svd}) as a function
  of rank $\kappa$. We see that the highest rank modes (corresponding
  to the smaller singular values) tend to contribute the most to the solution.\small }

\label{fig:contrs}
\end{center}
\end{figure}

\noindent
It is extremely informative to compare the amplitudes of the two terms
in the sum (\ref{eq:svd_sum}), and to monitor their contributions to
the solution vector with varying rank. In Figure \ref{fig:sing_values} we plot $|\bf{u}^T_\kappa
\cdot \bf{F}|$ and $\bf{u}_\kappa^T\bf{\cdot e}$ as a function
of rank $\kappa$, for a synthetic spectrum with a SNR per pixel of 50
(at a resolution of 3\AA) and
an exponentially-decaying star formation history. We observe the
behaviour described and discussed in \cite{OcvirkEtAl06}. The
combinations associated with the noise terms maintain a roughly
constant power across all ranks, with a an average value of
$\left\langle{\bf F}\right\rangle/$SNR. The data terms, however, drop significantly with rank,
and we can therefore identify two ranges: a noise-dominated $\kappa$-range,
in which the noise contributions match or dominate the true data
contributions, and a data-dominated range, where the contributions to
the solution are largely data motivated. We call the transition rank
$\kappa_{crit}$. Overall, high-$\kappa$ ranks tend
to dominate the solution, since the singular values $w_\kappa$ decrease with
$\kappa$. This only amplifies the problem by giving greater weight to
noise-dominated terms in the sum (\ref{eq:svd}). Figure
\ref{fig:contrs} shows the contribution coming from each rank $\kappa$
to the final solution - the coefficient $({\bf u}_\kappa^T \cdot {\bf F})/w_\kappa$. We see this weight increases with rank. 
\\
\par\noindent
Whereas this analysis gives us great insight into the problem, we do
not in fact use the sum (\ref{eq:svd}) to obtain ${\bf c}_{LS}$, for the
reasons given in section \ref{sec:bvls}. \\
\par\noindent
For real data we are only able to
calculate ${\bf u}_\kappa^T {\bf \cdot  F}$ and estimate the noise level at
$\left\langle{\bf
  F}\right\rangle/$SNR and we use this information to estimate the number of non-zero
parameters to recover from the data. Our aim is to a have a solution
which is dominated by the signal, and not by the noise. We therefore
want our number of non-zero recovered parameters to be less than or equal
to $\kappa_{crit}$. Estimating where this transition happens is always 
a noisy process. In this paper we take the conservative approach of
setting $\kappa_{crit}$ to be the rank at
which the perturbed singular values first cross the $\left\langle{\bf
  F}\right\rangle/$SNR barrier. In the case of Figure
\ref{fig:sing_values} this happens at $\kappa_{crit}=7$.

\subsection{Choosing a galaxy parametrization}
\label{sec:parametrization}

One of the advantages of VESPA is that it has the ability to choose
the number of parameters to recover in any given galaxy. This is
possible due to a time grid of varying resolutions, which VESPA can
explore to find a solution. This section describes this grid and the
criteria used to reach a final parametrization.\\

\subsubsection{The grid}
\label{sec:grid}
We work on a grid with a maximum resolution of $16$ age bins, logarithmically spaced in lookback time from 0.02 up to 14 Gyr. The grid has three
further resolution levels, where we split the age of
the Universe in eight, four and finally two age bins, also
logarithmically spaced in the same range.\\
\par\noindent
The idea behind the multi-resolution grid is to start our search with a
low number of parameters (in coarser resolution, so that the entirety
of the age of Universe is covered), and then increase the resolution
only where the data warrant it by splitting the bin with the highest
flux contribution in two, and so on. In effect, we construct one such
grid for each of the tabulated metallicities, $Z_a$ and $Z_b$. We work with five
metallicity values, $Z=[0.0004, 0.004, 0.008, 0.02, 0.05]$ which
correspond to the metallicity resolution of the models used,
where Z is the fraction of the mass of the star composed of metals ($Z_{\odot}=0.02$). 
The construction of the models for each of
the time bins is discussed in Section \ref{subsec:models}.\\
\par\noindent
To each of the grids we can apply a dust extinction as explained in
Section \ref{sec:dust}.

\subsubsection{The search}
\label{sec:search}
We go through the following steps in order to reach a solution:
\begin{enumerate}
\item We begin our search with three bins: two bins of width $4$ and one bin
  of width $8$ (oldest), where here we are measuring widths in units of high-resolution
  bins.
\item We calculate a solution using equation (\ref{eq:matrix}) for every
  possible combination of consecutive boundaries $Z_a$ and $Z_b$, and
  we choose the one which gives the best value of reduced $\chi^2$.
\item We calculate the number of perturbed singular values above the
  noise level, as described at the end of Section \ref{sec:noise_th}.
\item We find the bin which contributes the most to the total flux and
  we split it into two.
\item We find a solution in the new parametrization, this time by
  trying out all possible combinations of $Z_a$ and $Z_b$ for the
  newly split bins only, and fixing the metallicity boundaries of the
  remainder bins to the boundaries obtained in the previous
  solution. If a bin had no stars in the previous iteration, we set
  $Z_a=0.0004$ and $Z_b=0.05$.
\item We return to (iii) and we proceed until we have reached the
  maximum resolution in all populated bins. 
\item We look backwards in our sequence of solutions for the last instance
  with a number of non-zero recovered parameters equal to or less than
  $\kappa_{crit}$ as calculated in (iii) and take this as our best
  solution. 
\end{enumerate}
\par
\begin{figure*}
\begin{tabular}{l}
\includegraphics[width=0.95\linewidth]{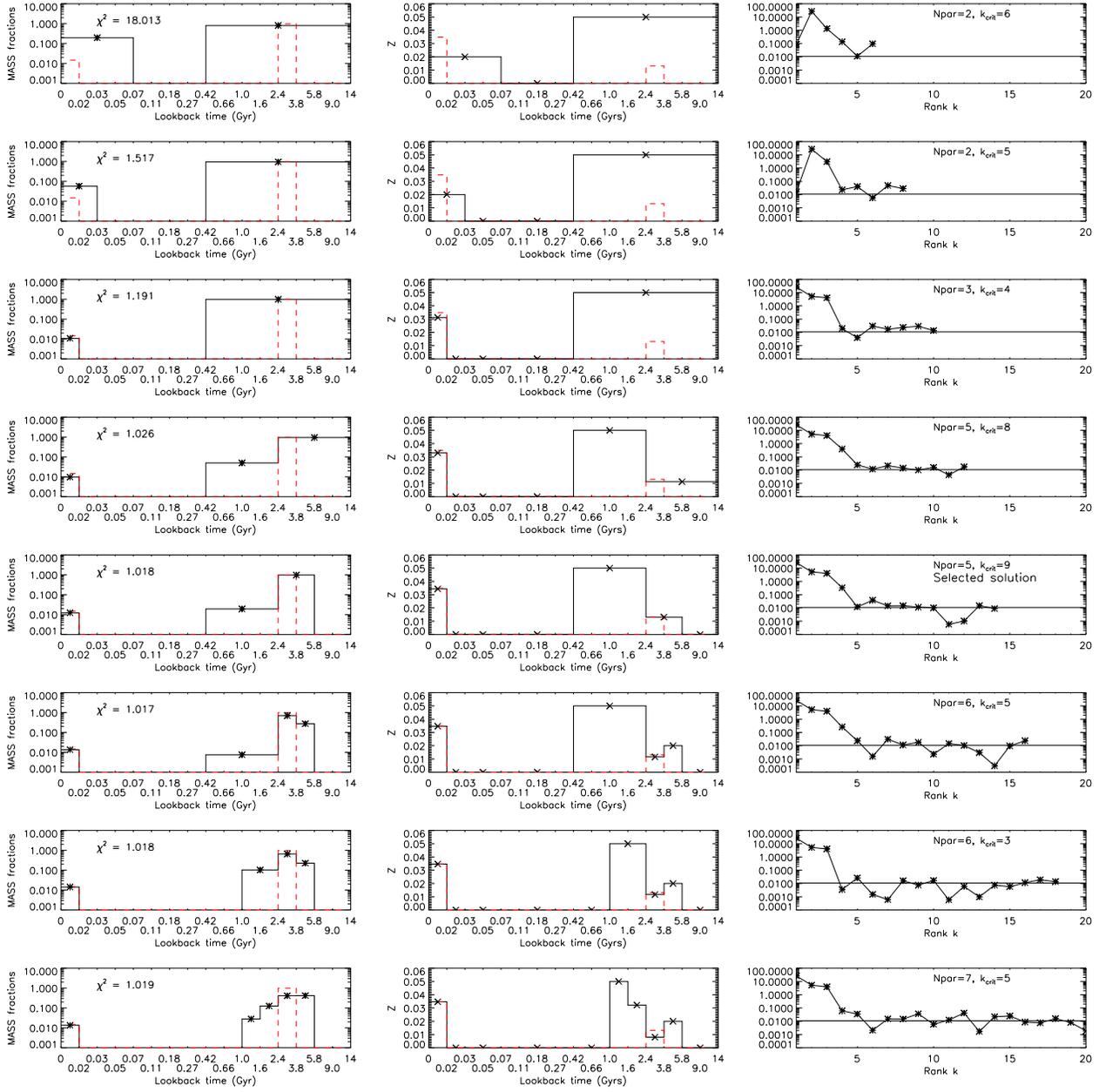}
\end{tabular}
\caption{The evolution of the fit, as VESPA searches for a
  solution. Sequence should be read from top to bottom. Each line
  shows a stage in the sequence: the left panel shows the input star formation history in the dashed
  line (red on the online version), and the recovered mass fractions
  on the solid line (black on the online version) for a given
  parametrization ; the middle panel shows the input metallicities in
  the dashed line (red on the online version), and the recovered
  metallicities on the solid line (black on the online version);
  the right panel shows the absolute value of the
  perturbed singular values $|\bf{u_\kappa \cdot F|}$ (stars and solid line)
  and the estimated noise level $\left\langle{\bf
  F}\right\rangle/$SNR. In this panel we also show the value of
  $\kappa_{crit}$ and the number of non-zero elements of ${\bf
  c_{LS}}$ in each iteration. The chosen solution is the {\bf fifth from the
  top}, and indicated
  accordingly. 
  This galaxy
  consists of 2 burst events of equal star
  formation rate - a very young and an old burst. It was modelled
  with a resolution of 3\AA\ and a
  signal-to-noise ratio per pixel of 50. We see the recovery is good but not perfect - there is a 1 per
  cent leakage from the older population - but better
  than the following solutions, where this bin is split. See text in
  Section \ref{sec:search} for more details. }
\label{fig:history}
\end{figure*}

\noindent
We illustrate this sequence in Figure \ref{fig:history}, where we show
the evolution of the search in a synthetic galaxy composed of two stellar
bursts of equal star formation rates - one young and one old. VESPA first
splits the components which contribute the most to the total flux. In
this case this is the young burst which can be seen in the first
bin. Even though VESPA always resolves bins with any mass to the
possible highest resolution, it then searches for the latest solution which
has passed the SVD criterion explained in Section \ref{sec:noise_th}. In
this case, this corresponds to the fifth from the top solution. VESPA
chooses this solution in favour of the following ones due to the number of
perturbed singular values above the solid line (right panel). In this
case, the solution chosen by VESPA is a better fit in parameters space
(note the logarithmic scale in the y-axis - the following
solution put the vast majority of the mass in the wrong bin). We
observed this type of improvement in the majority of all cases
studied (see Figure \ref{fig:svdtest}).    

\subsubsection{The final solution}
\label{sec:final_solution}

Our final solution comes in a parametrization such that the total number of non-zero
recovered parameters is less than or equal to the number of perturbed
singular values above the estimated noise level.\\
\par\noindent
The above sequence is performed for each of several combinations of
${\tau_V^{BC}, \tau_V^{ISM}}$, 
and we choose the attenuation which provides the best
fit. \\
\par\noindent
For each galaxy we recover N star formation masses, with an
associated metallicity, where N is the total number
of bins, and a maximum of two dust parameters.\\

\subsection{The models}
\label{subsec:models}
The backbone to our grid of models is the BC03
set of synthetic
SSP models \citep{BruzualEtCharlot03}, with a Chabrier initial mass
function \citep{Chabrier03} and Padova 1994 evolutionary tracks
\citep{AlongiEtAl93, BressanEtAl93, FagottoEtAl94a,
FagottoeEtAl94b, GirardiEtAl96} . Although any set of stellar population
models can be used, these provide a detailed spectral evolution of
stellar populations over a suitable range of wavelength, ages and
metallicities: $S(\lambda, t, Z)$. The models have been normalised to
one solar mass at the age $t=0$.

\subsubsection{High-resolution age bins}
\label{sec:HRbins}
At our highest resolution we work with 16 age bins, equally spaced in a logarithmic
time scale between now and the age of the Universe. In each bin, we assume a
constant star formation rate

\begin{equation}
f^{HR}_\alpha(\lambda, Z) = \psi \int_{\Delta t_\alpha}
S(\lambda, t, Z) dt
\end{equation}
\par\noindent
with $\psi = 1/\Delta t_\alpha$.

\subsubsection{Low-resolution age bins}

As described in Section \ref{sec:grid}, we work on a grid of different resolution time
bins and we construct the low resolution bins using the high
resolution bins described in Section \ref{sec:HRbins}. We do not assume a constant star
formation rate in this case, as in wider bins the light from the
younger components would largely dominate over the contribution from
the older ones. Instead, we use a decaying star formation history,
such that the light contributions from all the components are
comparable. Recall equation (\ref{eq:prob_int})
\begin{equation}
f^{LR}_\alpha(\lambda, Z) = \int_{\Delta t_\alpha} \psi(t) S(\lambda, t, Z) dt, 
\end{equation}
which we approximate to
\begin{equation}
f^{LR}_\beta(\lambda, Z) = \frac{\sum_{\alpha \in \beta} f^{HR}_\alpha(\lambda, Z) \psi_\alpha \Delta t_\alpha}{\sum_{\beta \in \alpha} \psi_\alpha \Delta t_\alpha}
\end{equation}
where low resolution bin $\beta$ incorporates the high resolution bins $\alpha \in
\beta$, and we set
\begin{equation}
\psi_\alpha \Delta t_\alpha = \frac{1}{\int_\lambda f^{HR}_\alpha(\lambda, Z)
  d\lambda}.
\label{eq:weights}
\end{equation}
\par\noindent
Depending on the galaxy, the final solution obtained with the sequence
detailed in Section \ref{sec:search} can be described in terms of
low-resolution age bins. In this case we should interpret the
recovered mass as the total mass formed during the period implied by
the width of the bin, but we cannot make any conclusions as to when in
the bin the mass was formed. Similarly, the recovered metallicity for
the bin should be interpreted as a mass-weighted metallicity for the
total mass formed in the bin.

\subsection{Errors}
The quality of our fits and of our solutions is affected by the noise
in the data, the noise in the models, and the parametrization we
choose (which does not reflect the complete physical scenario within a
galaxy). We aim to apply VESPA firstly to SDSS galaxies, which
typically have a SNR $\approx 20$ per resolution element of $3$\AA, which puts us in a regime where the main
limitations come from the noise in the data.\\
\par\noindent
To estimate how much noise affects our recovered solutions we take a rather
empirical approach. For each recovered solution we create $n_{error}$
random noisy realisations and we apply VESPA to each of these
spectra. We re-bin each recovered solution in the parametrization of
the solution we want to analyse and estimate the
covariance matrices
\begin{equation}
\label{eq:cx}
C(x)_{\alpha \beta} = \left\langle (x_\alpha - \bar{x}_\alpha)(x_\beta - \bar{x}_\beta)
\right\rangle
\end{equation}
\begin{equation}
C(Z)_{\alpha \beta} = \left\langle (Z_\alpha - \bar{Z}_\alpha)(Z_\beta - \bar{Z}_\beta) \right\rangle.
\end{equation}
All the plots in Sections \ref{sec:tests} and \ref{sec:results} show
error bars derived from $C_{\alpha \alpha}^{1/2}$, although it is worth keeping
in mind that these are typically highly correlated.

\subsection{Timings}
A basic run of VESPA (which consists of roughly 5 runs down the
sequence detailed in Section \ref{sec:search}, one for each value of
dust extinction) takes about 5
seconds. If accurate error estimations are needed per galaxy, this
will add another one or two minutes to the timing, depending on how
accurately one would like to estimate the covariance matrices, and 
depending on the 
number of data points. With $n_{error}=10$, a typical SDSS galaxy takes around one minute
to analyse.


\section{Tests on Simulated Data}
\label{sec:tests}
\begin{figure*}
\begin{tabular}{ll}
\includegraphics[width=0.5\linewidth,clip]{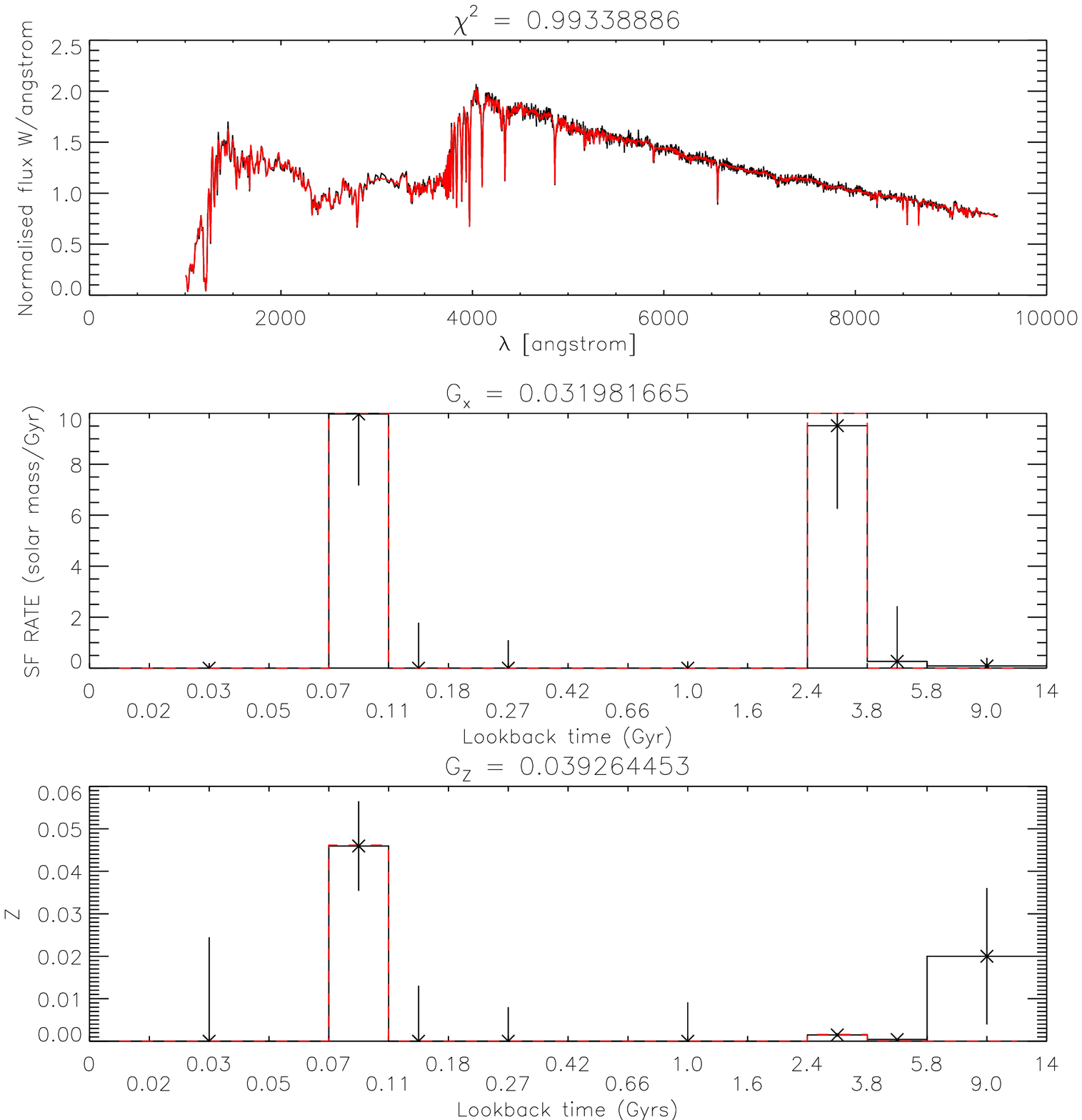}
\includegraphics[width=0.5\linewidth,clip]{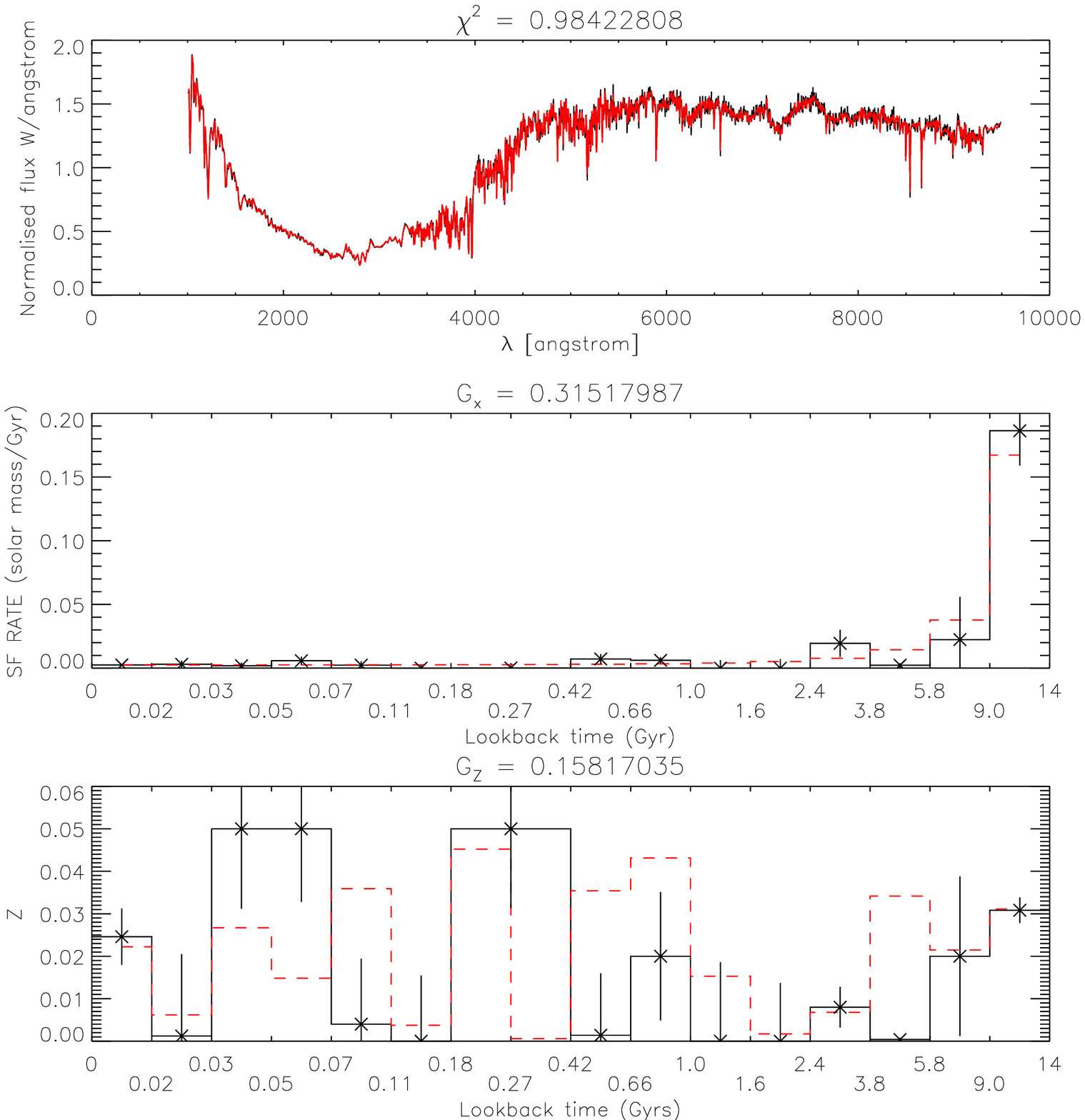}
\end{tabular}
\caption{Two examples of VESPA's analysis on synthetic galaxies. The
  top panels show the original spectrum in the dark line (black in the
  online version) and fitted spectrum in the lighter line (red in the
  online version ). The
  middle panels show the input (dashed, red) and the recovered (solid,
  black) star formation rates and the bottom panel shows the input
  (dashed, red) and recovered (solid, black) metallicities per
  bin. Note that even though many of the recovered metallicities are
  wrong, these tend to correspond to bins with very little star
  formation, and are therefore virtually unconstrained. 
}
\label{fig:SFH_ex}
\end{figure*}
\begin{figure*}
\begin{tabular}{ll}
\includegraphics[width=0.32\linewidth,clip]{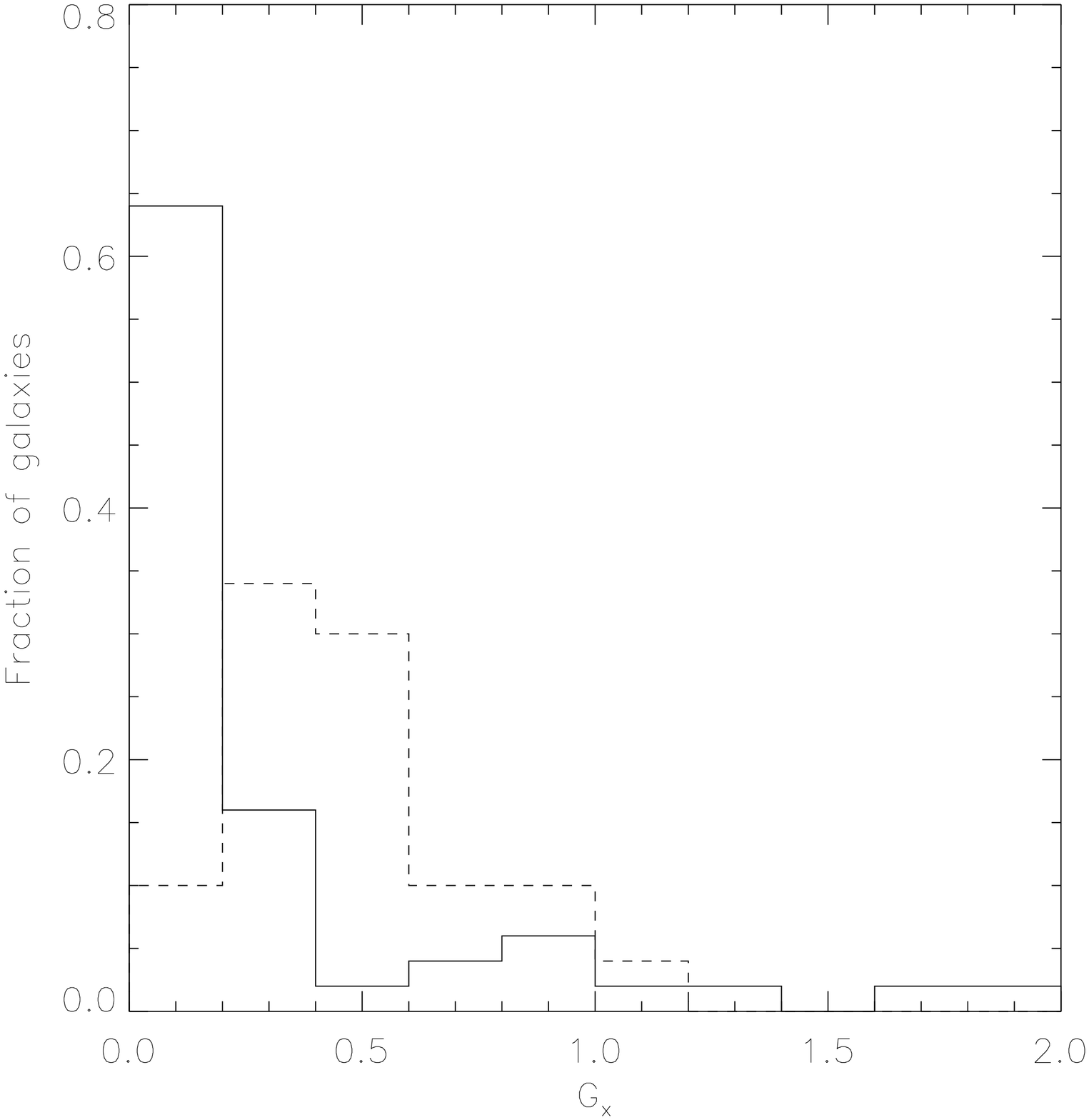}
\includegraphics[width=0.32\linewidth,clip]{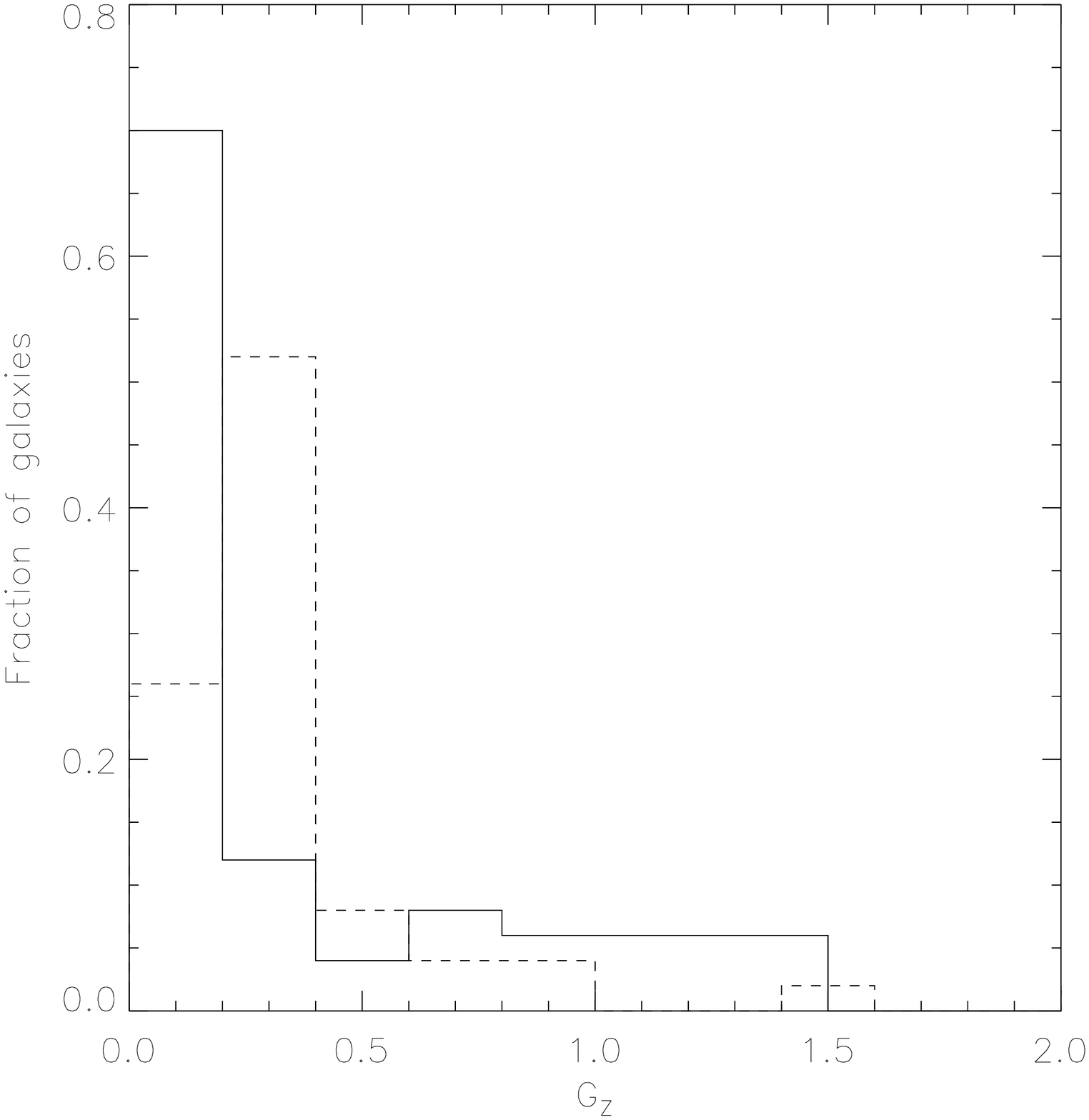}
\includegraphics[width=0.32\linewidth,clip]{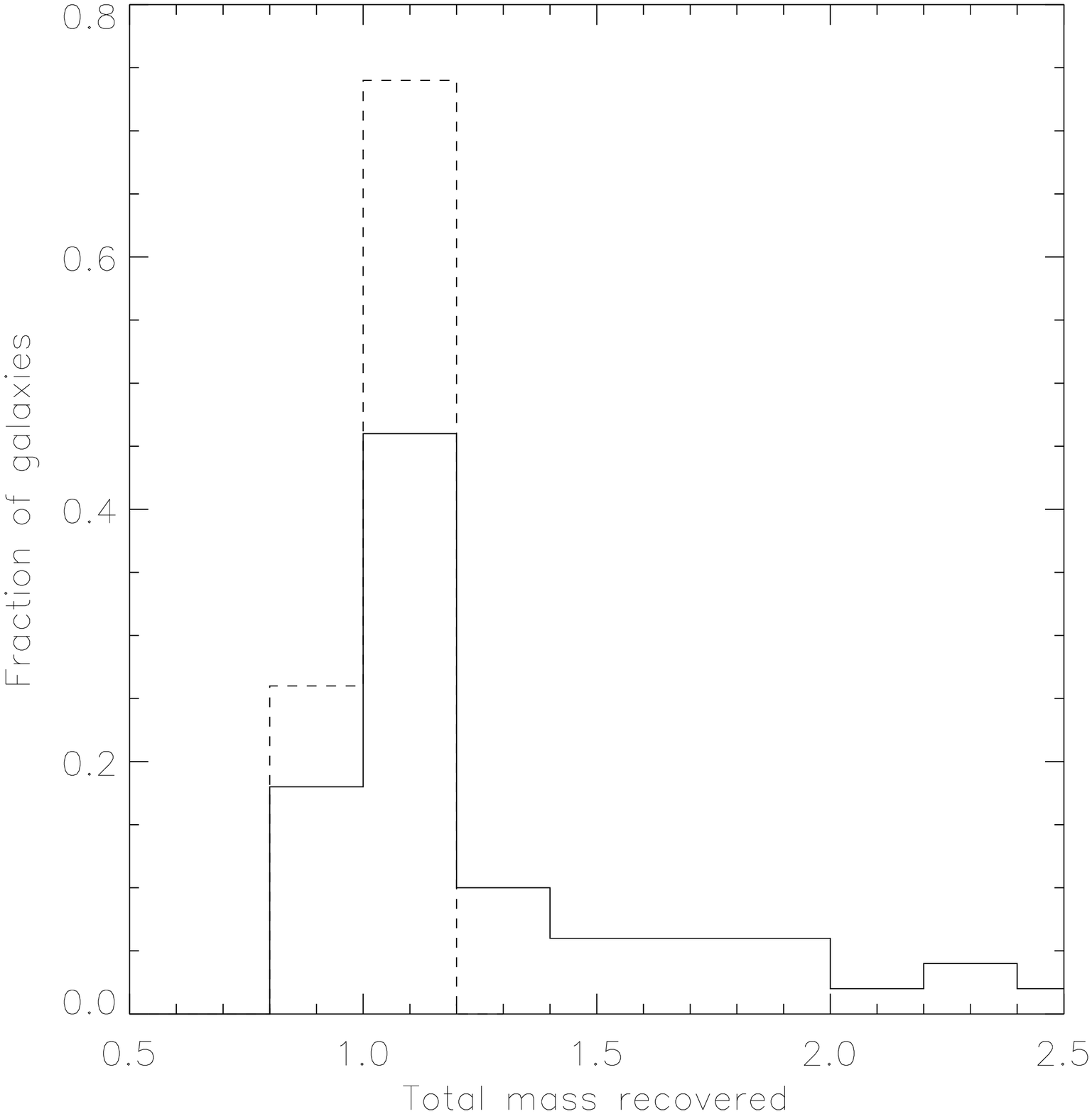}
\end{tabular}
\caption{The distribution of $G_x$, $G_Z$ and total mass recovered for
  50 galaxies with a
  SNR per pixel of 50. Solid lines correspond to dual
  burst and dashed lines to exponentially decaying
  ones. See
  text in Section \ref{sec:SFH} for details.
}
\label{fig:G_dual50_long}
\end{figure*}

We tested VESPA on a variety of synthetic spectra, in order to
understand its capabilities and limitations. In particular, we tried
to understand the effect of three factors in the quality of our
solutions: the input star formation history, the noise in the
data, and the wavelength coverage of the spectrum. We have also looked
at the effects of dust extinction. Throughout we have
modelled our galaxies in a resolution of 3\AA.\\
\par\noindent
Even though we are aware that showing individual examples of VESPA's
results from synthetic spectra can be extraordinarily unrepresentative, we
feel obliged to show a few for illustration purposes. We
will show a typical result for most of the cases we present,
but we also define some measurements of success, so that the overall
performance of VESPA can be tracked as we vary any factors. We define

\begin{equation}
G_x=\sum_\alpha \left|\frac{x_\alpha - x^I_\alpha}{x^I_\alpha}\right| \omega_\alpha
\end{equation}
and
\begin{equation}
G_Z=\sum_\alpha\left|\frac{Z_\alpha - Z_\alpha^I}{Z^I_\alpha}\right| \omega_\alpha
\end{equation}
\par\noindent
where $x_\alpha^I$ and $Z_\alpha^I$ are the total mass and correspondent
metallicity in bin $\alpha$
(re-binned to match the solution's parametrization if necessary), and
$\omega_\alpha$ is the flux contribution of population of age $t_\alpha$. $G_x$ and
$G_Z$ are a flux-weighted average of the total absolute fractional
errors in the solution, and give an indication of how well VESPA
recovers the most significant parameters. A perfect solution gives
$G_x = G_Z = 0$. It is also worth noting that this statistic does not
take into account the associated error with each recovered parameter -
deviations from the true solution are usually expected given the
estimated covariance matrices. 
We will also show how these factors affect the recovered total mass
for a galaxy. In all cases we have re-normalised the total masses such that
total input mass for each galaxy is 1.
\subsection{Star formation histories}
\label{sec:SFH}
\begin{figure*}
\begin{tabular}{ll}
\includegraphics[width=0.5\linewidth,clip]{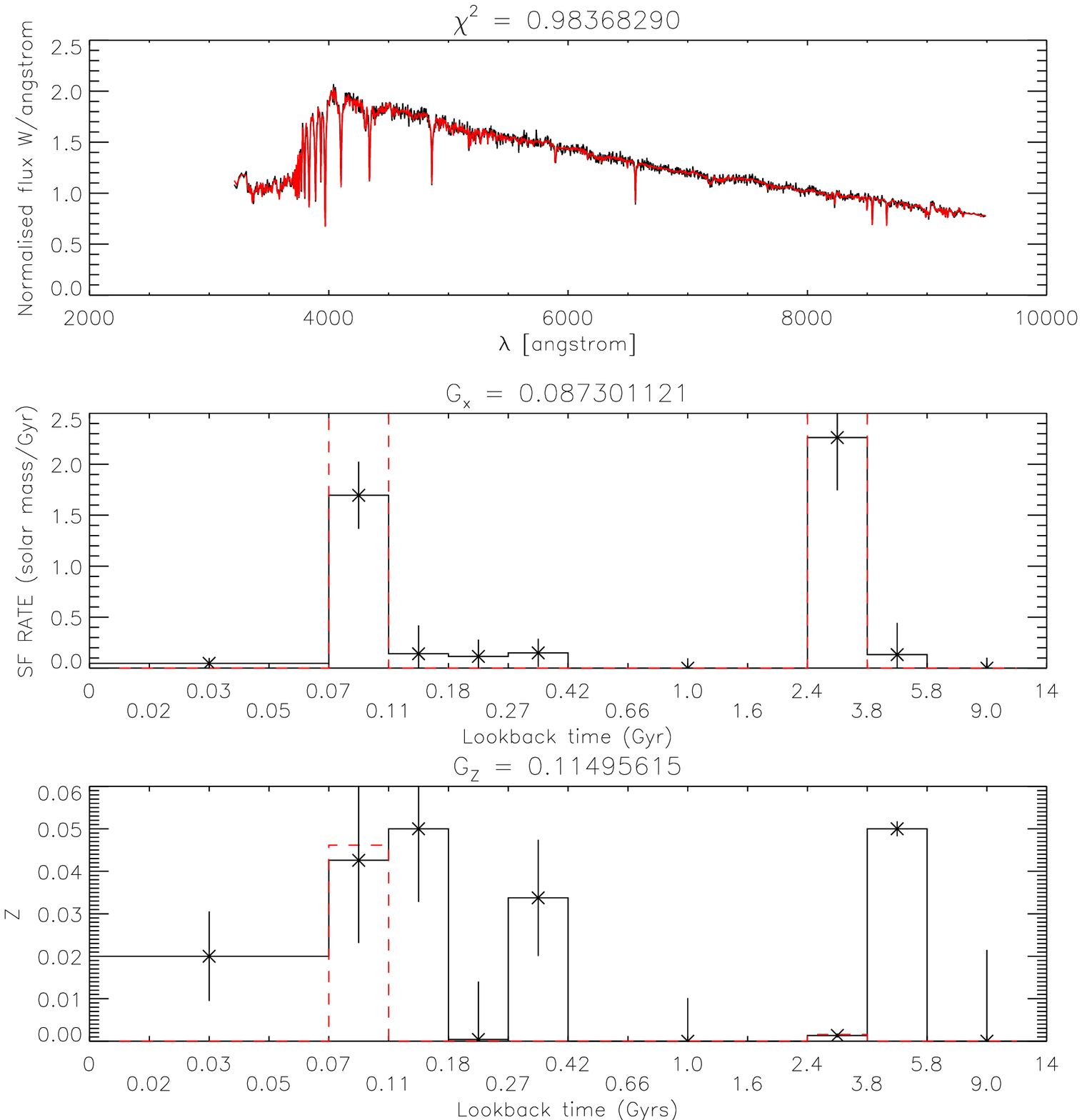}
\includegraphics[width=0.51\linewidth,clip]{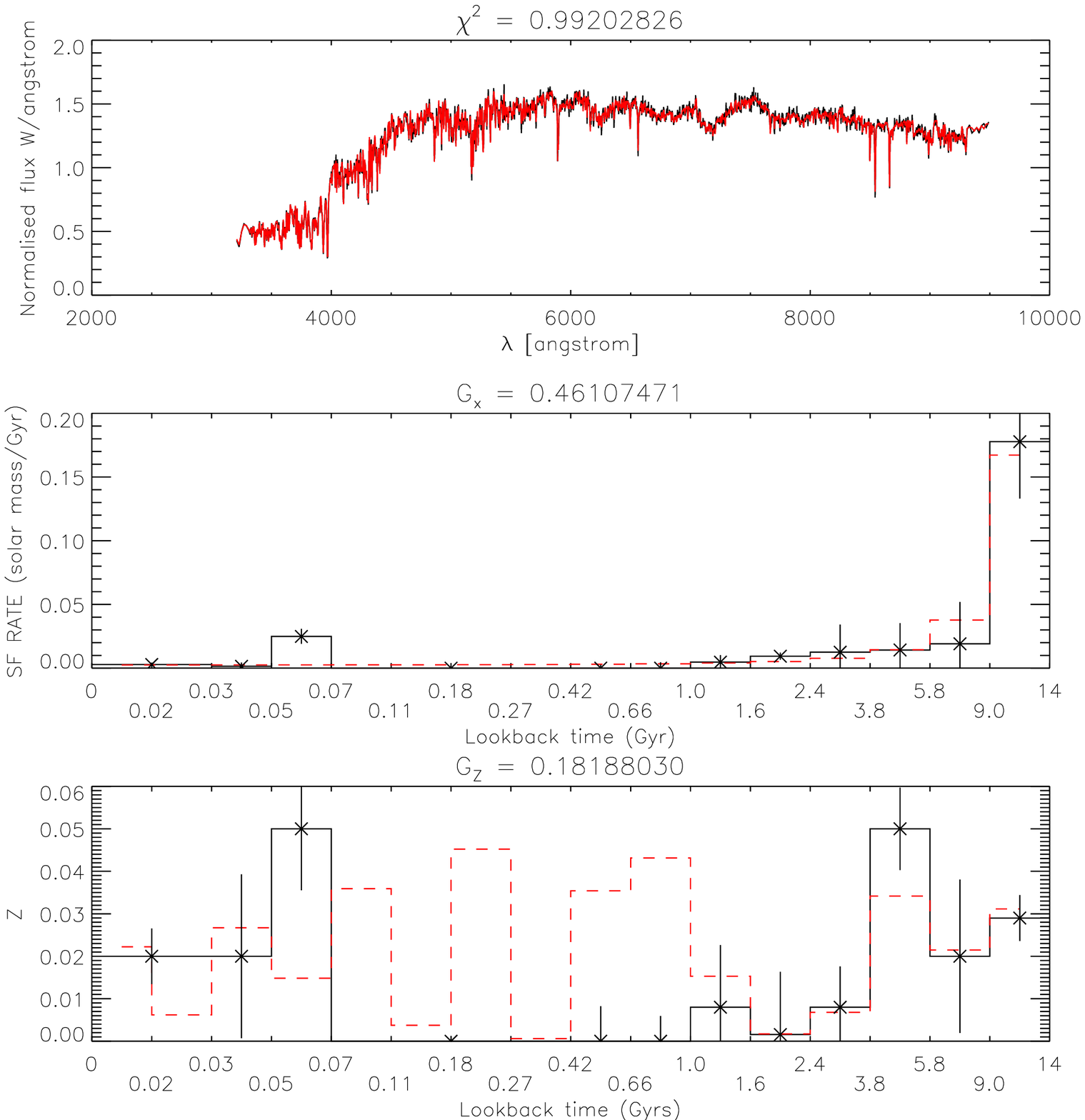}
\end{tabular}
\caption{Same galaxies as in Figure \ref{fig:SFH_ex}, but results
  obtained by using a smaller wavelength range. The goodness-of-fit in
  data space is still excellent, but it becomes more difficult to break
  certain degeneracies. 
}
\label{fig:lambda_ex}
\end{figure*}
\par\noindent
\begin{figure*}
\begin{tabular}{ll}
\includegraphics[width=0.32\linewidth,clip]{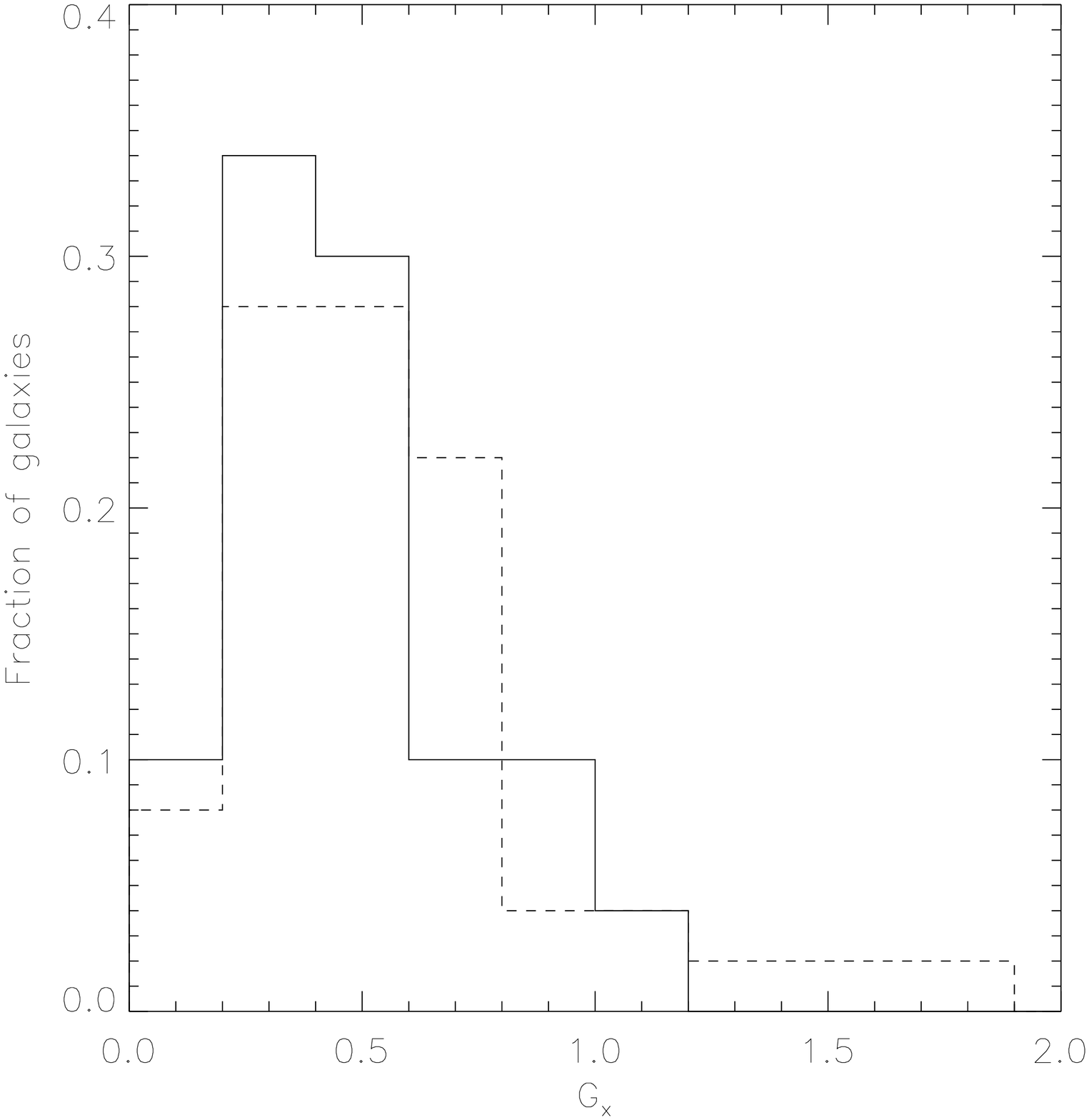}
\includegraphics[width=0.32\linewidth,clip]{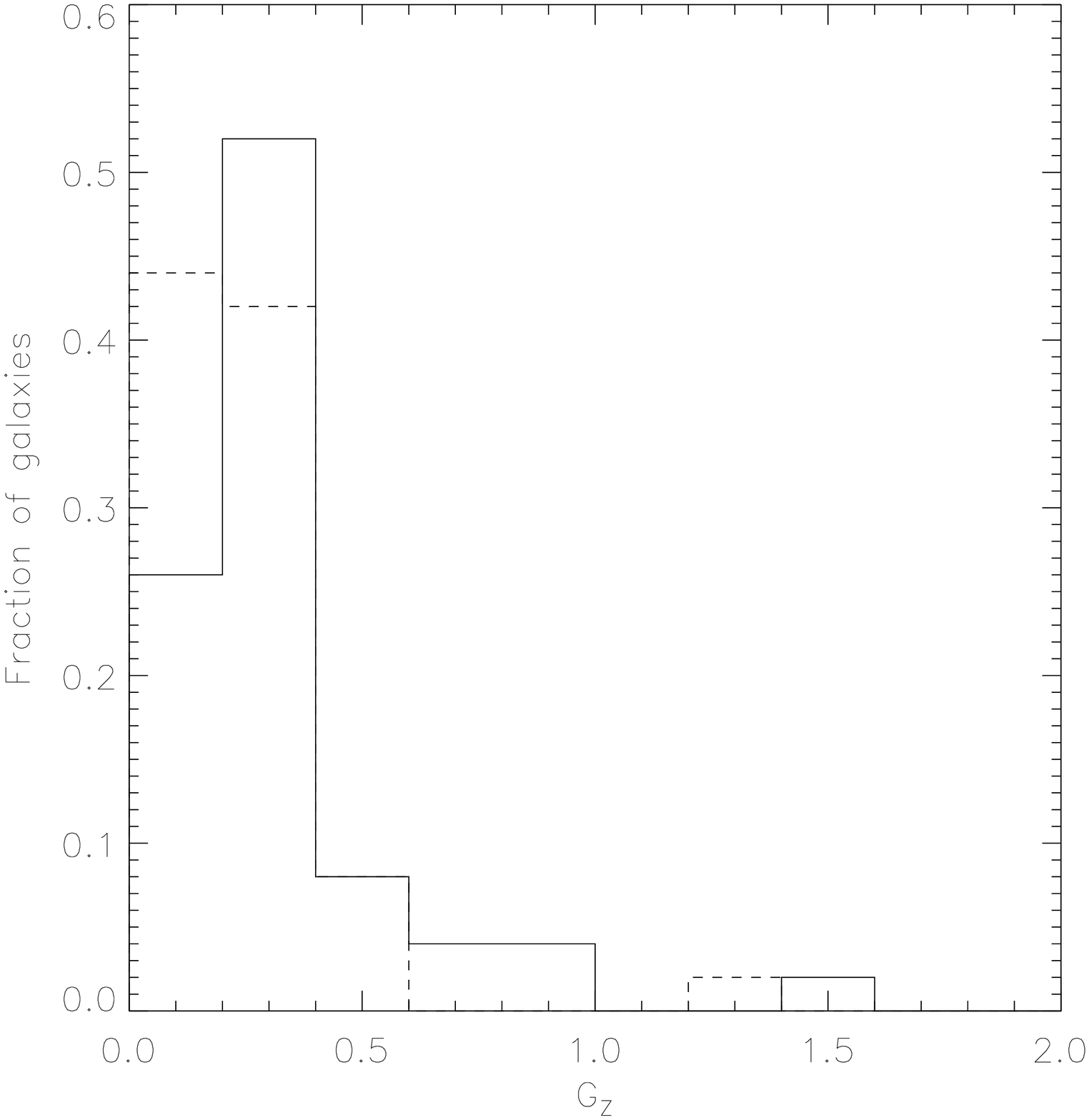}
\includegraphics[width=0.32\linewidth,clip]{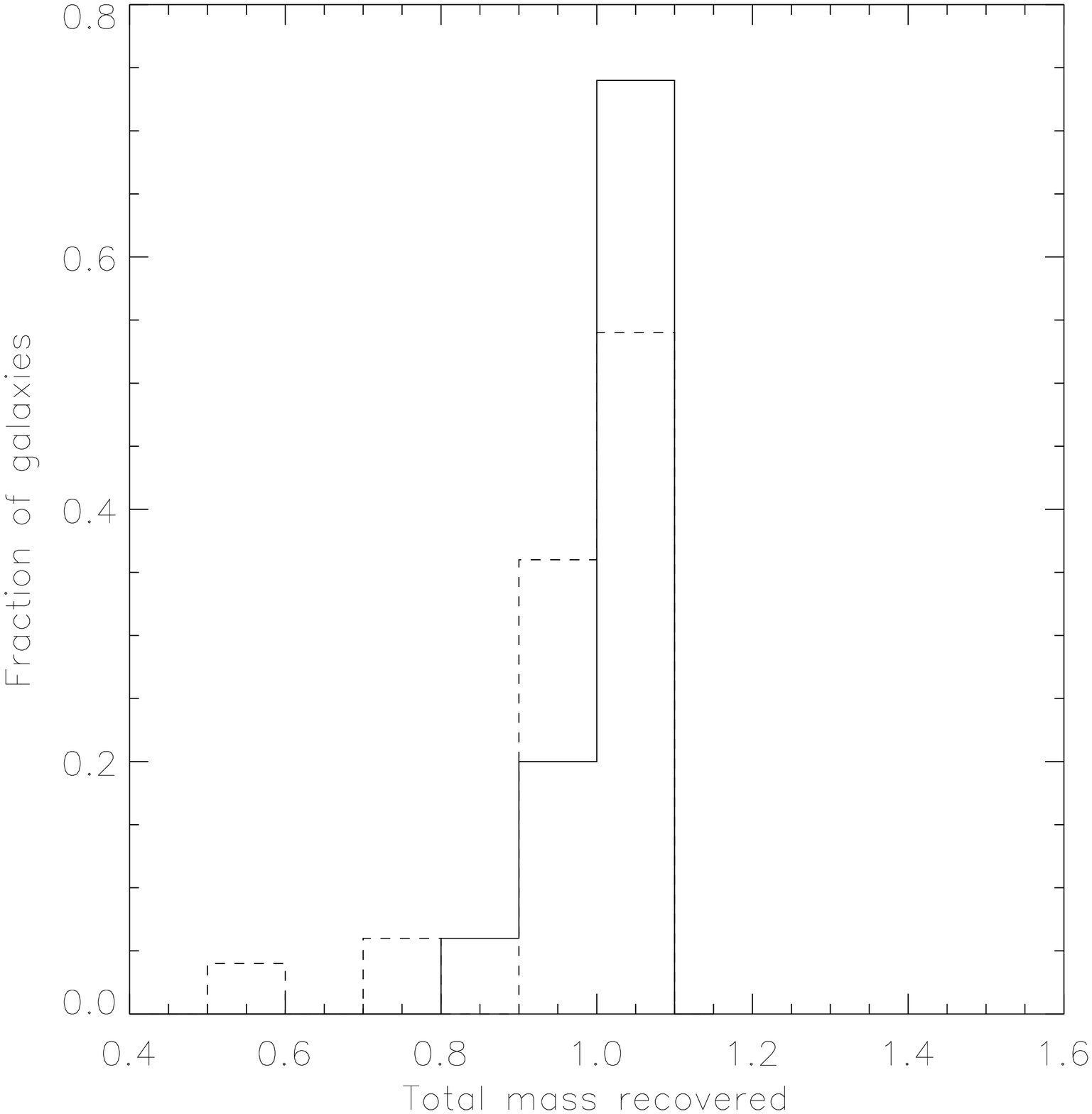}
\end{tabular}
\caption{The distribution of $G_x$, $G_Z$ and total mass recovered
  for 50 galaxies with a
  SNR per pixel of 50 and two different wavelength coverage. Solid
  line corresponds to $\lambda \in [1000,9500]\AA$ and dashed line to
  $\lambda \in [3200,9500]$ \AA.
}
\label{fig:G_lambda}
\end{figure*}
\par\noindent
We present here some results for synthetic spectra with two different
star formation histories. All of the spectra in this section were
synthesised with a SNR per pixel of 50, and we initially fit the very
wide wavelength
range $\lambda \in [1000,9500]$\AA.  \\
\par\noindent
We choose two very difference cases: firstly a star formation history of
dual bursts, with a large random variety
of burst age separations and metallicities (where we set the star
formation rate to be 10 solar masses per Gyr in all bursts). Secondly, we chose a SFH
with an
exponentially decaying star formation rate: SFR $\propto \exp(\gamma
t_\alpha)$, where $t_\alpha$ is the age of the bin in lookback time in Gyr. Here we
show results for $\gamma = 0.3$ Gyr$^{-1}$. Rather than being physically
motivated, our choice of
$\gamma$ reflects a SFH which is not too steep as to essentially mimic a
single
old burst, but which is also not completely dominated by recent star
formation. In all cases the metallicity in each bin is randomly
set. Figure \ref{fig:SFH_ex} shows a typical example from each
type. \\
\par\noindent
Figure \ref{fig:G_dual50_long} shows the distribution of $G_x$, $G_Z$
and of the recovered total masses for a sample of 50 galaxies.
We see differences between the two cases. Firstly, in dual bursts
galaxies, we seem to do better in recovering data from significant individual
bins, but worse in overall mass. This reflects the fact that $G_x$ is
dominated by the fractional errors in the most significant bins, but
the total mass can be affected by small flux contributions in old
bins which can have large masses. On the other hand, with an
exponentially decaying star formation rate, we do worse overall (although this is mainly a reflection that more bins have
significant contributions to the flux) but we recover the total
mass of the galaxy exceptionally well. 

\subsection{Wavelength range}
\begin{figure}
\vspace{0.6cm}
\begin{center}
\includegraphics[width=3.in]{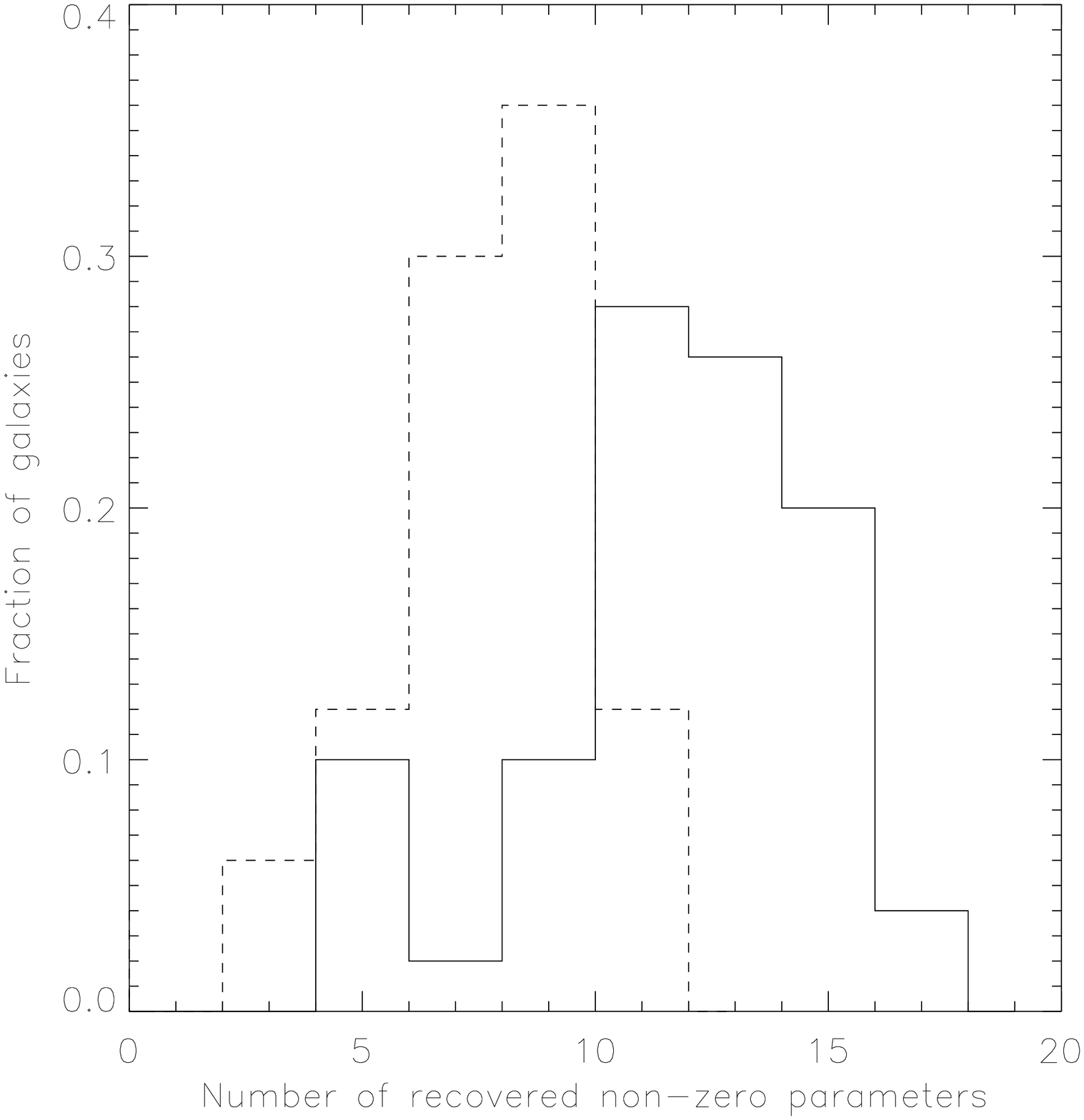}
\caption{The recovered number of non-zero parameters in 50 galaxies
  with an exponentially decaying star formation history, using different wavelength ranges: $\lambda \in
  [1000,9500]$ \AA (solid line) and $\lambda \in [3200, 9500]$ \AA (dashed
  line). Please note that these correspond to the {\it total} number of non-zero
  components in the solution vector $\bf{c_\kappa}$ and not to the number
  of recovered stellar populations. }
\label{fig:Npar_dual}
\end{center}
\end{figure}

Wavelength range is an important factor in this sort of analysis, as
different parts of the spectrum will help to break different
degeneracies. Since we are primarily interested in SDSS galaxies, we
have studied how well VESPA does in the more realistic wavelength
range of $\lambda \in [3200,9500]$ \AA. \\
\par\noindent
Figure \ref{fig:lambda_ex} shows the results for the same galaxies
shown in Figure \ref{fig:SFH_ex}, obtained with the new wavelength
range. In these particular cases, we notice a more pronounced
difference in the dual bursts galaxy, but looking at a more substantial sample of
galaxies shows that this is not generally the case. Figure
\ref{fig:G_lambda} shows $G_x$, $G_Z$ and total mass recovered for 50
exponentially decaying star formation history galaxies, with a signal
to noise ratio of 50 and the two different wavelength ranges. We do
not see a largely significant change in both cases, and we observe 
a less significant difference in the dual bursts galaxies (not plotted
here).\\
\par\noindent
We find it instructive to keep track of how many
parameters we recover in total, as we change any factors. Figure
\ref{fig:Npar_dual} shows an histogram of the total number of non-zero
parameters we recovered from our sample galaxies with
exponentially-decaying star formation histories and both wavelength
ranges. Note that these are the components of the solution vector $\bf{c_\kappa}$
which are non-zero - they do not represent a number of recovered
stellar populations. In this
case there is a clear decrease in the number of recovered parameters,
suggesting a wider wavelength range is a useful way to increase
resolution in parameter space.\\
\par\noindent
\subsection{Noise}
\label{sec:noise}
\begin{figure*}
\begin{tabular}{ll}
\includegraphics[width=0.32\linewidth,clip]{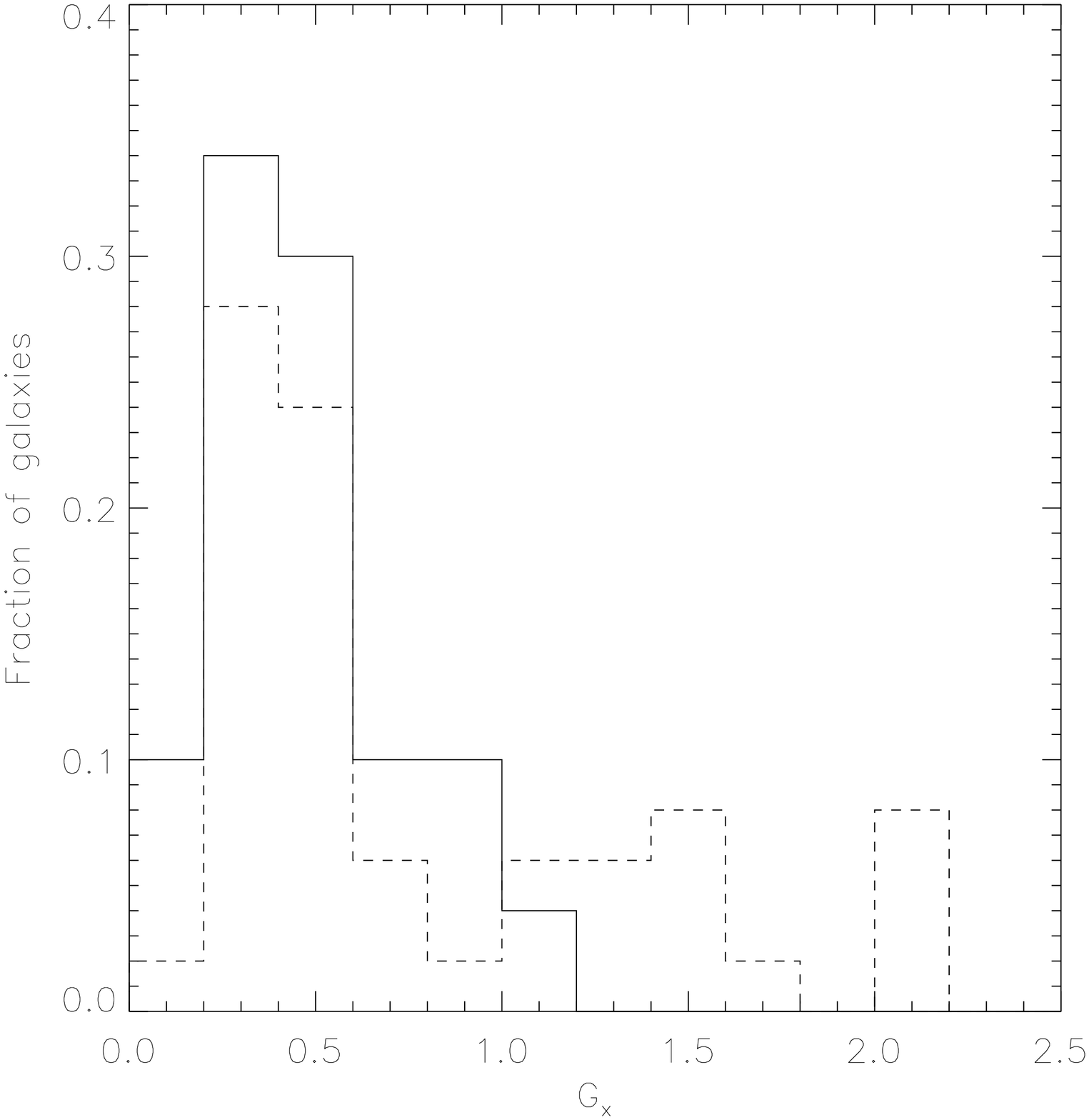}
\includegraphics[width=0.32\linewidth,clip]{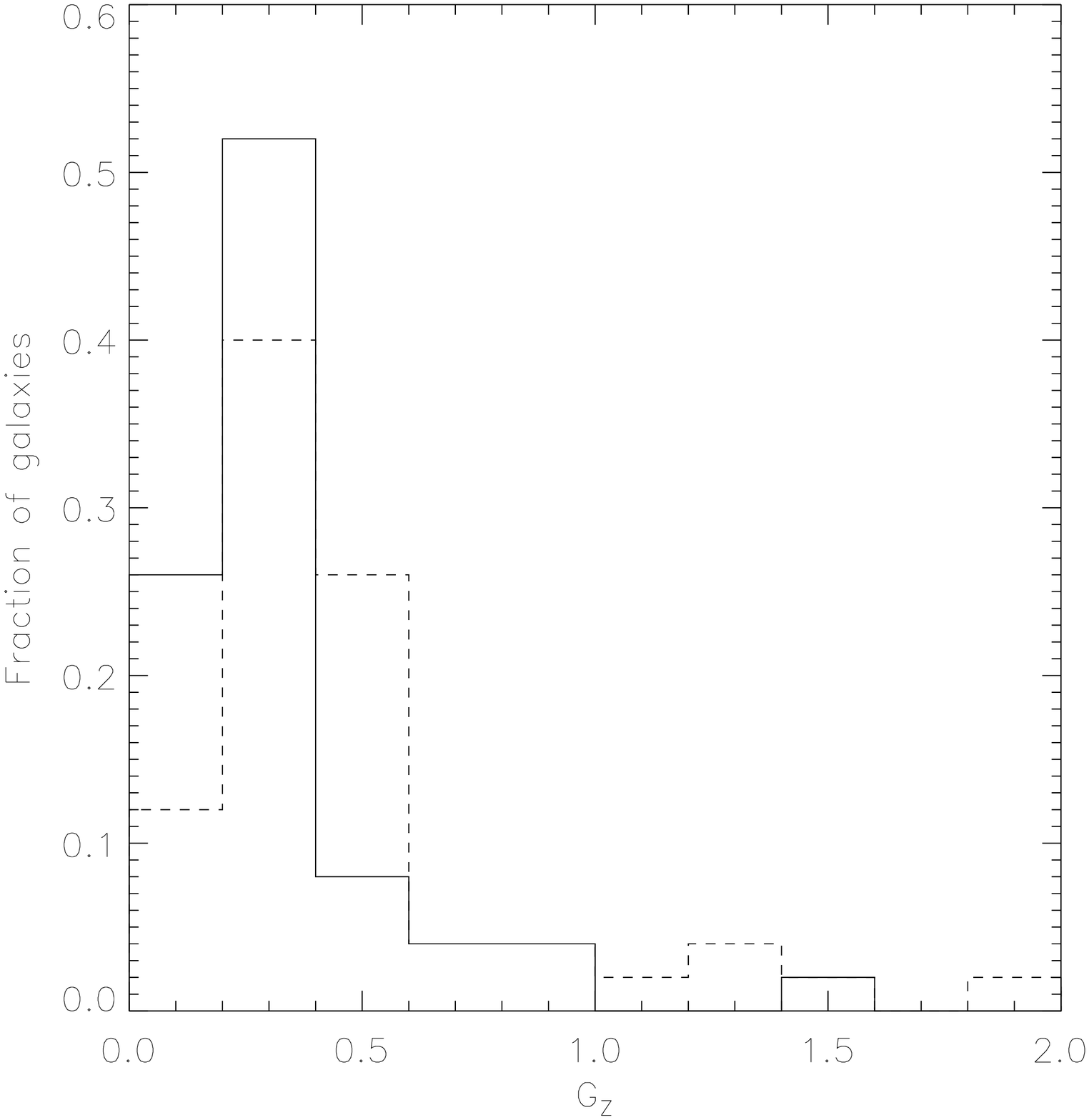}
\includegraphics[width=0.32\linewidth,clip]{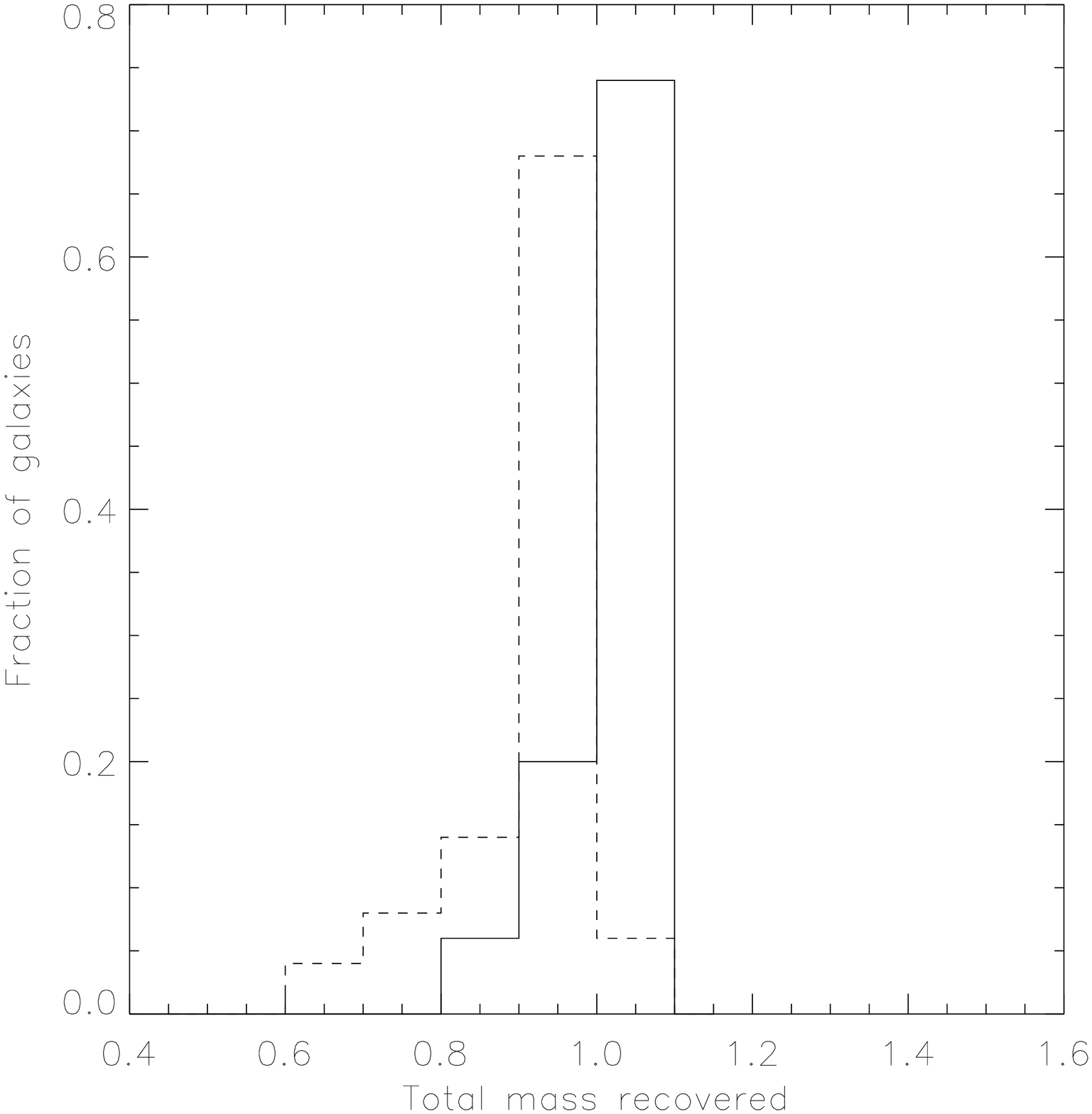}
\end{tabular}
\caption{The distribution of $G_x$, $G_Z$ and total mass recovered
  for 50 galaxies with an exponential decaying star formation history and
  different signal-to-noise ratios. Solid lines correspond
  to SNR $= 50$ and dashed lines to SNR =$20$. See
  text in Section \ref{sec:noise} for details.
}
\label{fig:G_SNR}
\end{figure*}

It is of interest to vary the signal-to-noise ratio in the
synthetic spectra. We have repeated the studies detailed in the two
previous sections with varying values of noise, and we investigate how
this affects both the quality of the solutions and their resolution in
parameter space. \\
\par
\begin{figure}
\vspace{0.6cm}
\begin{center}
\includegraphics[width=3.in]{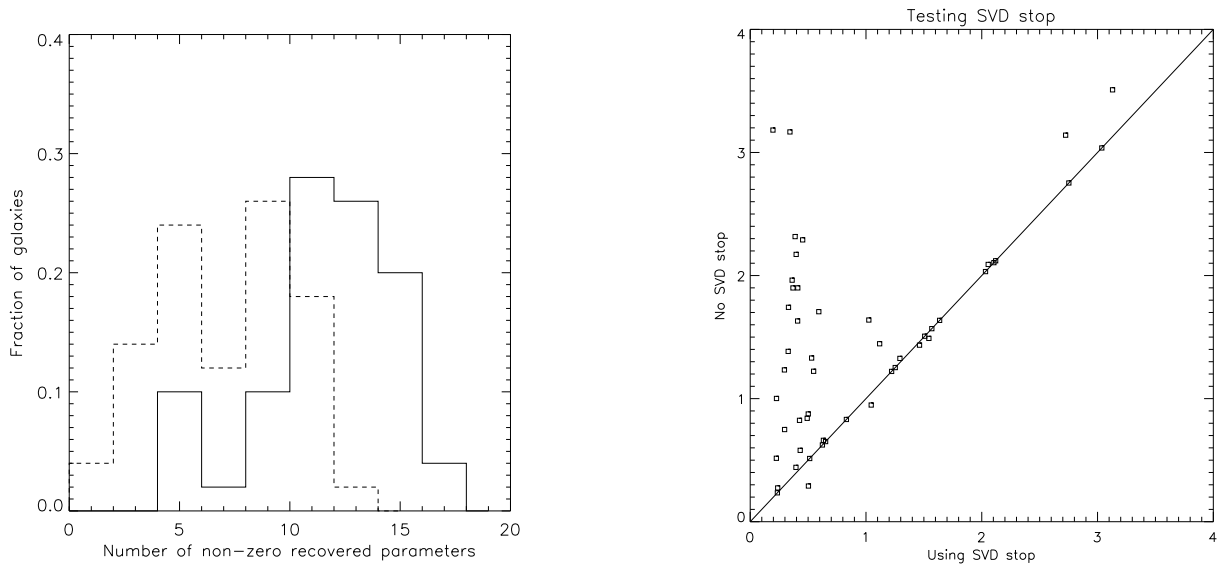}
\caption{The recovered number of non-zero parameters as we change the
  noise in the data from 50 (solid line) to 20 (dashed line), in a
  sample of galaxies with an exponentially decaying star formation
  rate. Please note that these correspond to the {\it total} number of non-zero
  components in the solution vector $\bf{c_\kappa}$ and not to the number
  of recovered stellar populations.}
\label{fig:Npar_exp}
\end{center}
\end{figure}

\noindent
Figure \ref{fig:Npar_exp} shows how the recovered number of parameters
changed by increasing the noise in the galaxies with an exponentially
decaying star formation rate and wide wavelength range.
In this case the increase in the noise leads to a
significant reduction of the number of parameters recovered for each
galaxy. This behaviour is equally clear for different star
formation histories and different wavelength coverage, and is
directly caused by the stopping criterion defined in Section \ref{sec:noise}. \\
\par
\begin{figure}
\vspace{0.6cm}
\begin{center}
\includegraphics[width=3.in]{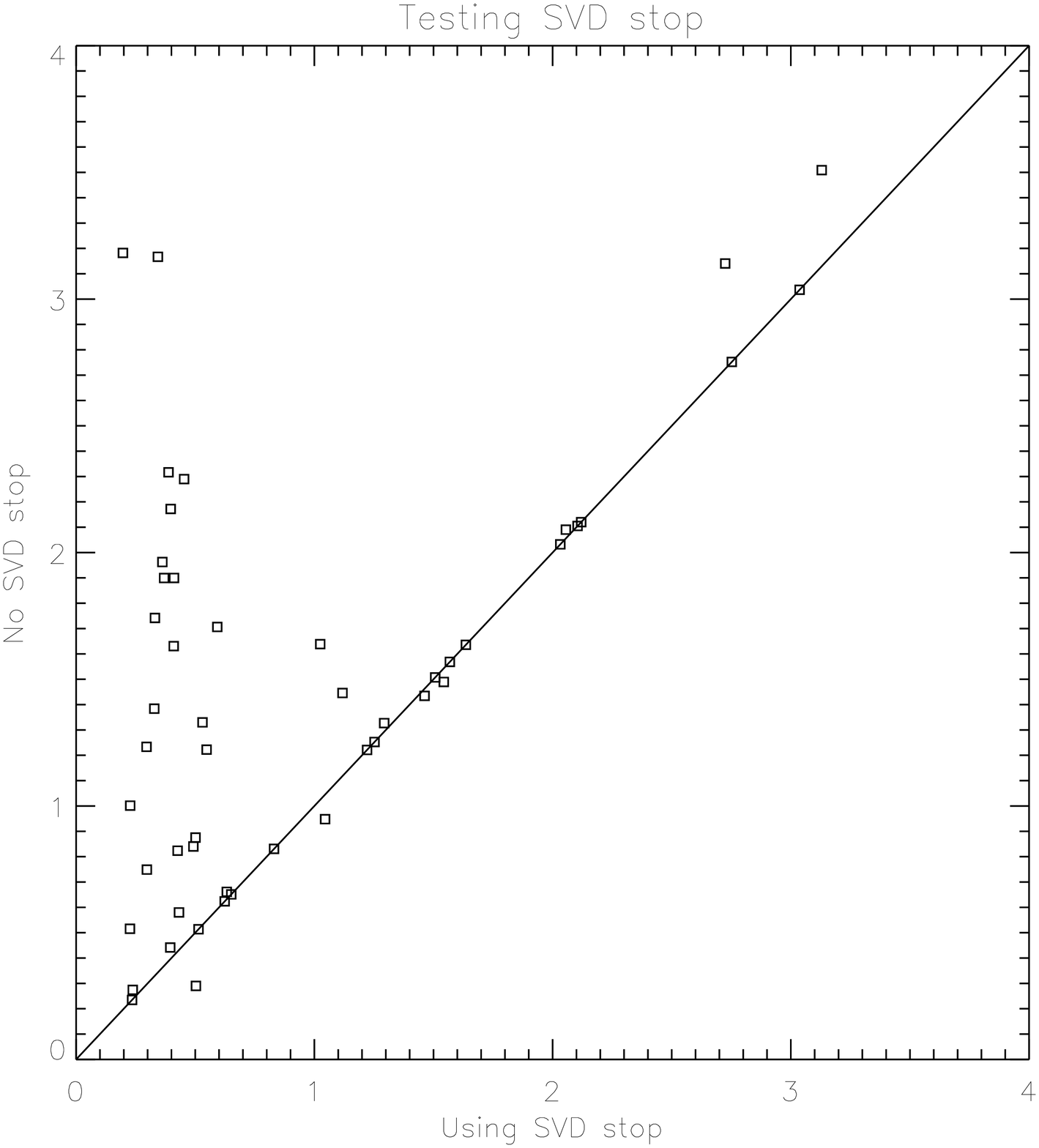}
\caption{Testing the SVD stopping criterion. Plots show goodness of fit $G_x$ for
the solution of 50 galaxies obtained with and without the SVD stopping
criterion. We see that by recovering only as many parameters as the
data warrants gives improved parameter estimation in almost all cases,
and a striking improvement in many. }
\label{fig:svdtest}
\end{center}
\end{figure}

\noindent
The quality of the solutions is also affected by this increase in
noise, as can be seen in figure \ref{fig:G_SNR}, where we have plotted
$G_x$, $G_Z$ and the total recovered mass for
two different values of SNR. The quality of the solutions decreases
with the higher noise levels, as is to be expected.
However, a more interesting question to ask is
whether this decrease in the quality of the solutions would indeed be
more pronounced without the SVD stopping criterion. Figure \ref{fig:svdtest} shows a
comparison between $G_x$ obtained as we have described and obtained
without any stopping mechanism (so letting our search go to the
highest possible resolution and taking the final solution) for 50
galaxies with an exponentially decaying star formation history and a
signal-to-noise ratio of 20. The results show clearly that there is a
significant advantage in using the SVD stopping criterion. Naturally, the
goodness of fit in data space is consistently better as we increase
the number of parameters but this improvement is illusory - the
parameter recovery is worse. This is exactly the expected behaviour - we
choose to sacrifice resolution in parameter space in favour of a more
robust solution - even though naively one could think a lower $\chi^2$
solution would indicate a better solution. The significance of this
improvement changes with the amount of noise and wavelength range of
the data (and to a lesser
extent with type of star formation history) but we observed an 
improvement in all cases we have studied. \\
\par\noindent
As expected, further decreasing the signal-to-noise ratio leads to a further
degradation of the recovered solutions. This is
accompanied by a suitable increase in the error bars and correlation
matrices, but in cases of a SNR$\approx 10$ and less it becomes very
difficult to recover any meaningful information from individual
spectra. 

\subsection{Dust}
\label{sec:dust_synth}
In this section we use simulated galaxies to study the effect of
dust in our solutions. As explained in Section \ref{sec:dust}, due
to the non-linear nature of the problem, we cannot include dust as
one of the free parameters analysed by SVD. Instead, we fit for a maximum of two
dust parameters using a brute-force approach which aims to minimise
$\chi^2$ in data-space by trying out a series of values for
$\tau_V^{ISM}$ and $\tau_V^{BC}$. \\
\par\noindent
For each galaxy we assign random values of
$\tau_V^{ISM} \in [0,2]$ and $\tau_V^{BC} \in [1,2]$ and we are
interested in how well we recover these parameters and any possible degeneracies.\\
\par
\begin{figure}
\vspace{0.6cm}
\begin{center}
  \includegraphics[width=3.in]{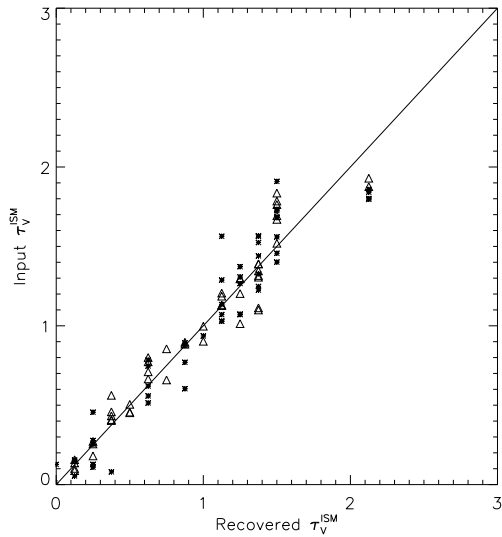}
\caption{Testing the recovery of $\tau_V^{ISM}$ for 50 galaxies with a
  exponentially-decaying star formation history (triangles) and 50
  galaxies formed with a random combination of dual bursts
  (stars). The input values are randomly chosen and continuously distributed
  between 0 and 2. The recovered values are chosen from a tabulated
  grid between 0 and 4.
}
\label{fig:dusttest}
\end{center}
\end{figure}

\noindent
Figure \ref{fig:dusttest} shows the input and recovered values
for $\tau_V^{ISM}$ for galaxies with a signal-to-noise ratio of 50, and which were
analysed using the wavelength range $\lambda \in [3200, 9500]$\AA. We
show results for two different cases of star formation history: 50
galaxies with an exponentially-decaying SFR and 50 galaxies formed by
dual-bursts. We observe a good recovery of $\tau_V^{ISM}$ in both
cases, especially at low optical depths. \\
\par\noindent
However, we mostly observe a poor recovery of $\tau_V^{BC}$,
especially at high optical depths. This is unsurprisingly flagging up
a certain level of degeneracy between mass and degree of extinction,
which gets worse as the optical depth increases. Essentially, it
becomes difficult to distinguish between a highly obscured massive
population and a less massive population surrounded by less dust.
It is
worth keeping in mind that young populations are affected by both dust
components simultaneously, and generally, even though
the recovery of the second dust parameter may not be accurate, it
allows for a better estimation of the dominant dust component. \\
\par\noindent
This can be tested by
simulating galaxies on a two-component dust model and
by analysing them using both a single component model, and a
two-component model. E.g., when using the more sophisticated model, we
noted that the mean error on $\tau_V^{ISM}$ on a subsample of
dual-burst galaxies (synthesised as explained in section
\ref{sec:SFH}, but chosen to have young star formation) was reduced from 35 to 28 per cent.
This simple test also revealed that we are less
likely to underestimate the mass of young populations by allowing an
extra dust component, but that we are also introducing an extra
degeneracy, especially so in the case of faint young
populations. However, we feel that the two-parameter dust model brings
more advantages than disadvantages, with the caveat being that dusty young
populations can be poorly constrained. In any case, we note that each
galaxy is always analysed with a one-parameter model before being
potentially analysed with a two-parameter model, and both solutions
are kept and always available for analysis. \\
\par\noindent
Finally, our test also partly justifies our choice to first
run a single dust component model and only apply a two-component model
if we detect stars in the first two bins - we find that although a
one-component model might underestimate the amount of young stars, it
does not fail to detect them. We repeated a similar test in real data,
by analysing the same sample with and one- and a two-parameter dust
model. We found similar results, with a one-parameter model failing to
yield star formation in young bins only around 1 per-cent of the time
(compared to the two-parameter model), and only in cases where the
contribution of the light from the young populations was very small
(of the order of 1 to 2 per cent).


\section{Results}
\label{sec:results}

In this section we present some results obtained by applying VESPA to
galaxies in the SDSS. Our aim is to analyse
these galaxies, and to produce and publish a catalogue of robust star formation histories,
from which a wealth of information can then be derived. We leave this
for another publication, but we present here results from a sub-sample
of galaxies, which we used to test VESPA in a variety of ways. 

\subsection{Handling SDSS data}

Prior to any analysis, we processed the SDSS spectroscopic data, so as
to accomplish the desired spectral resolution and
mask out any wanted signal.\\
\par\noindent
The SDSS data-files supply a mask vector, which flags any potential
problems with the measured signal on a pixel-by-pixel basis. We use
this mask to remove any unwanted regions and emission lines. In
practical terms, we ignore any pixel for which the provided mask value
is not zero. \\
\par\noindent
The BC03 synthetic models produce outputs at a resolution of 3\AA,
which we convolve with a Gaussian velocity dispersion curve with a
stellar velocity $\sigma_V= 170$km$s^{-1}$, this being a typical value
for SDSS galaxies. We
take the models' tabulated wavelength values as a fixed grid and
re-bin the SDSS data into this grid, using an inverse-variance weighted
average. We compute the new error vector accordingly. Note that the
number of SDSS data points averaged into any new bin is not constant,
and that the re-binning process is done after we have masked out any
unwanted pixels. Additionally to the lines yielded by the mask vector,
we mask out the following emission line regions in every spectrum's rest-frame
wavelength range: [5885-5900,
  6702-6732, 6716-6746, 6548-6578, 6535-6565, 6569-6599, 4944-4974,
  4992-5022, 4846-4876, 4325-4355, 4087-4117, 3711-3741, 7800-11000] \AA. \\
\par\noindent
These re-binned data- and noise-vectors are essentially the ones we use in our
analysis. However, since the linear algebra assumes white-noise, we pre-whiten the data and construct a new
flux vector $F'_{j} = F_{j}/\sigma_j$, which has unit variance, $\sigma'_j = 1, \forall
j$, and a new model matrix $A'_{ij} = A_{ij}/\sigma_j$.

\subsection{Duplicate galaxies}
\label{sec:duplicates}
There are a number of galaxies in the SDSS database which have been
observed more than once, for a variety of reasons. This provides an
opportunity to check how variations in observation-dependent corrections affect the
results obtained by VESPA. \\
\par\noindent
We have used a subset of the sample of duplicate
objects in \cite{BrinchmannEtal04}\footnote{Available at 
http://www.mpa-garching.mpg.de/SDSS/} to create two sets of
oservations for 2000 galaxies, which we
named list A and list B. We are interested in seeing how the errors we
estimate for our results compare to errors introduced by intrinsic
variations caused by changing the observation conditions (such as quality
of the spectra, placement of the fibre, sky subtraction or
spectrophometric calibrations). \\
\par
\begin{figure}
\includegraphics[width=3.0in]{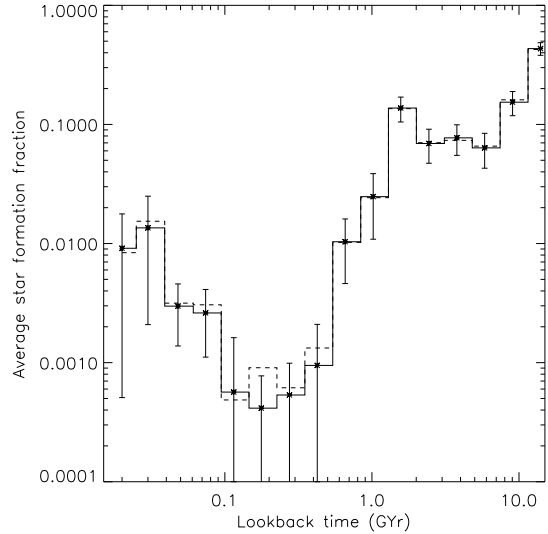}
\caption{Average star formation fraction as a function of lookback
  time for the 2000 galaxies in list A (solid line) and list B (dashed
  line). The error bars shown are the errors bars on the mean for each
  age bin. We show only the errors from list A to avoid cluttering -
  the errors from list B are of similar amplitude.
}
\label{fig:duplicates_sff}
\end{figure}

\noindent
Figure \ref{fig:duplicates_sff} shows the average star formation
fraction as a function of lookback time for both sets of
observation. The error bars showed are errors on the mean. We see no
signs of being dominated by systematics when estimating the star
formation fraction of a sample of galaxies. \\
\par\noindent
\par
\begin{figure}
\includegraphics[width=3.0in]{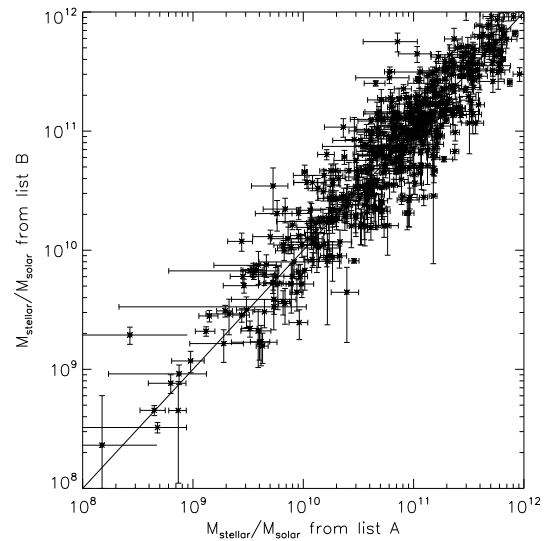}
\caption{Total stellar mass recovered for two sets of observations of
  500 galaxies in the main galaxy sample. The error bars are calculated from $C(x)$.
}
\label{fig:duplicates_masses}
\end{figure}

\noindent
Figure \ref{fig:duplicates_masses}
shows the total stellar mass obtained for a set of 500 galaxies in both
observations (details of how
we estimate the total stellar mass of a galaxy are included in section
\ref{sec:vespamoped}). The error bars are obtained 
directly from the estimated covariance matrix $C(x)$ (equation
\ref{eq:cx}). Even though most of the galaxy duplicates produce mass estimates
in agreement with each other given the error estimates, a minority
does not. Upon inspection, these galaxies show significant differences
in their continuum, but after further investigation it remains unclear what motivates such a
difference. The simplest explanation is that the spectrophotometric
calibration differs significantly between both observations, and that
might have been the reason the plate or object was
re-observed. Whatever the reason however, the clear conclusion is
that stellar mass estimates are highly sensitive to changes in the spectrum
continuum, and the errors we estimate from the covariance matrix alone
might be too small. \\
\par\noindent
\par\noindent
We did not find any signs of a systematic bias in any of the analysis
we carried out. \\

\subsection{Real fits}
In this section we discuss the quality of the fits to SDSS galaxies
obtained with VESPA. \\
\par
\begin{figure}
\includegraphics[width=3in]{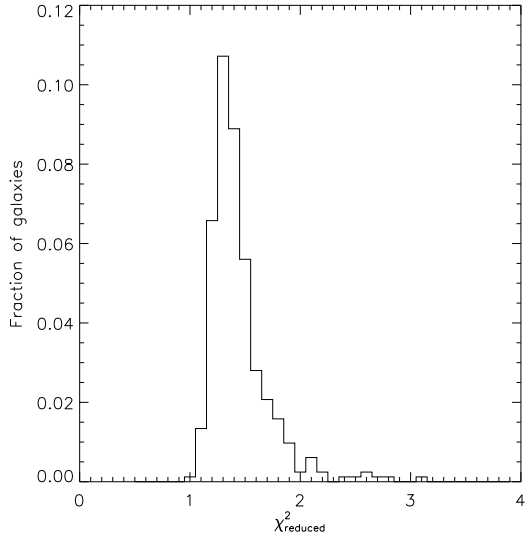}
\caption{The distribution of reduced values of $\chi^2$ for a sample
  of 360 galaxies analysed by VESPA.
}
\label{fig:chi2}
\end{figure}

\begin{figure*}
\begin{tabular}{l}
\includegraphics[width=6.0in]{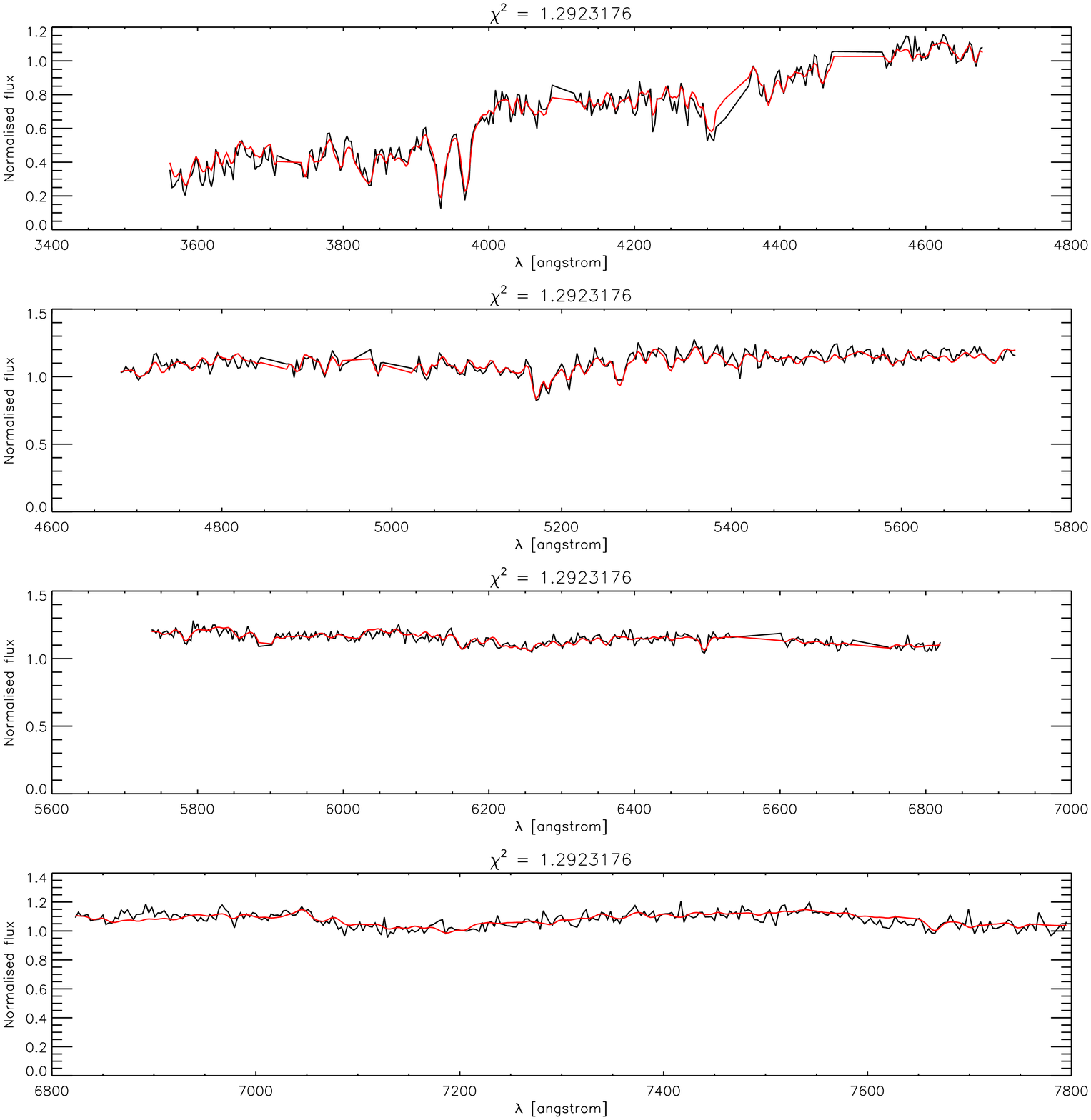}
\end{tabular}
\caption{Typical fit to a galaxy from the SDSS. The dark line is the real
  data (arbitrary normalisation), and the lighter line (red on the online version) is VESPA's
  fit to the data.  
}
\label{fig:fit}
\end{figure*}

\noindent
As explained in Section 2, VESPA finds the best
fit solution in a $\chi^2$ sense for a given parametrisation, which is
self-regulated in order to not allow an excessive number of fitting
parameters. We have shown that this self-regularization gives a better
solution in parameter space (Figure \ref{fig:svdtest}), despite often not allowing
the parametrization which would yield the best fit in data
space (Figure \ref{fig:history}). However, our aim is still to find a solution which gives a good
fit to the real spectrum. Figure \ref{fig:chi2} shows the 1-point
distribution of reduced values of reduced $\chi^2$ for 1 plate of
galaxies. This distribution peaks at around $\chi^2_{reduced} = 1.3$,
and figure \ref{fig:fit} shows a fit to one of the galaxies with a
typical value of goodness of fit. \\
\par\noindent
It is worth noting that the majority of the
fits which are most pleasing to the eye, correspond to the ones with a
high signal to noise ratio and high value of reduced
$\chi^2$. One would expect the best fits to come from the
galaxies with the best signal. However, we believe the fact that they
do not is not a limitation of the method, but a limitation of the
modelling. There are a number of reasons why VESPA would be unable to produce
very good fits to the SDSS data. One is the adoption of a single
velocity 
dispersion (170 kms$^{-1}$) which could easily be improved upon at the expense of CPU time. 
However, the dominant reason is likely to be lack of accuracy in
stellar and dust modelling - whereas BC03 models can and do reproduce
a lot of the observed features, it is also well known that this sucess
is limited as there are certain spectral features not yet accurately
modelled, or even modelled at all. There are similar deficiencies in
dust models and dust extinction curves. The effect of
the choice of modelling should not be overlooked, and we refer the
reader to a discussion in Section 4.5 of \cite{PanterEtAl06}), where
these issues are discussed.  \\


\subsection{VESPA and MOPED}
\label{sec:vespamoped}
In this Section we take the opportunity to compare the results from
VESPA and MOPED, obtained from the same sample of galaxies. The VESPA
solutions used here are obtained with a one-parameter dust
model, to allow a more fair comparison between the two
methods. Both methods make similar assumptions regarding stellar
models, but MOPED uses an LMC \citep{GordonEtAl03} dust extinction curve, and single
screen modelling for all optical depths. \\
\par\noindent
Our sample consist of two plates from the SDSS DR3
\citep{Abazajian05MNRAS} 
(plates 0288 and 0444),
from which we analyse a total of 821 galaxies. We are mainly
interested in comparing the results in a global sense.
MOPED in its standard configuration attempts to recover 23 parameters
(11 star formation fractions, 11 metallicities and 1 dust parameter),
so we might expect considerable degeneracies. Indeed, in the past the
authors of MOPED
have cautioned against using it to interpret individual galaxy
spectra too precisely. We have observed degeneracies between adjacent
bins in MOPED, but on the other hand a typical MOPED solution has many
star formation fractions which are essentially zero, so the number of
significant contributions is always much less than 23. 

\par\noindent
\begin{figure}
\begin{tabular}{ll}
\includegraphics[width=3.0in]{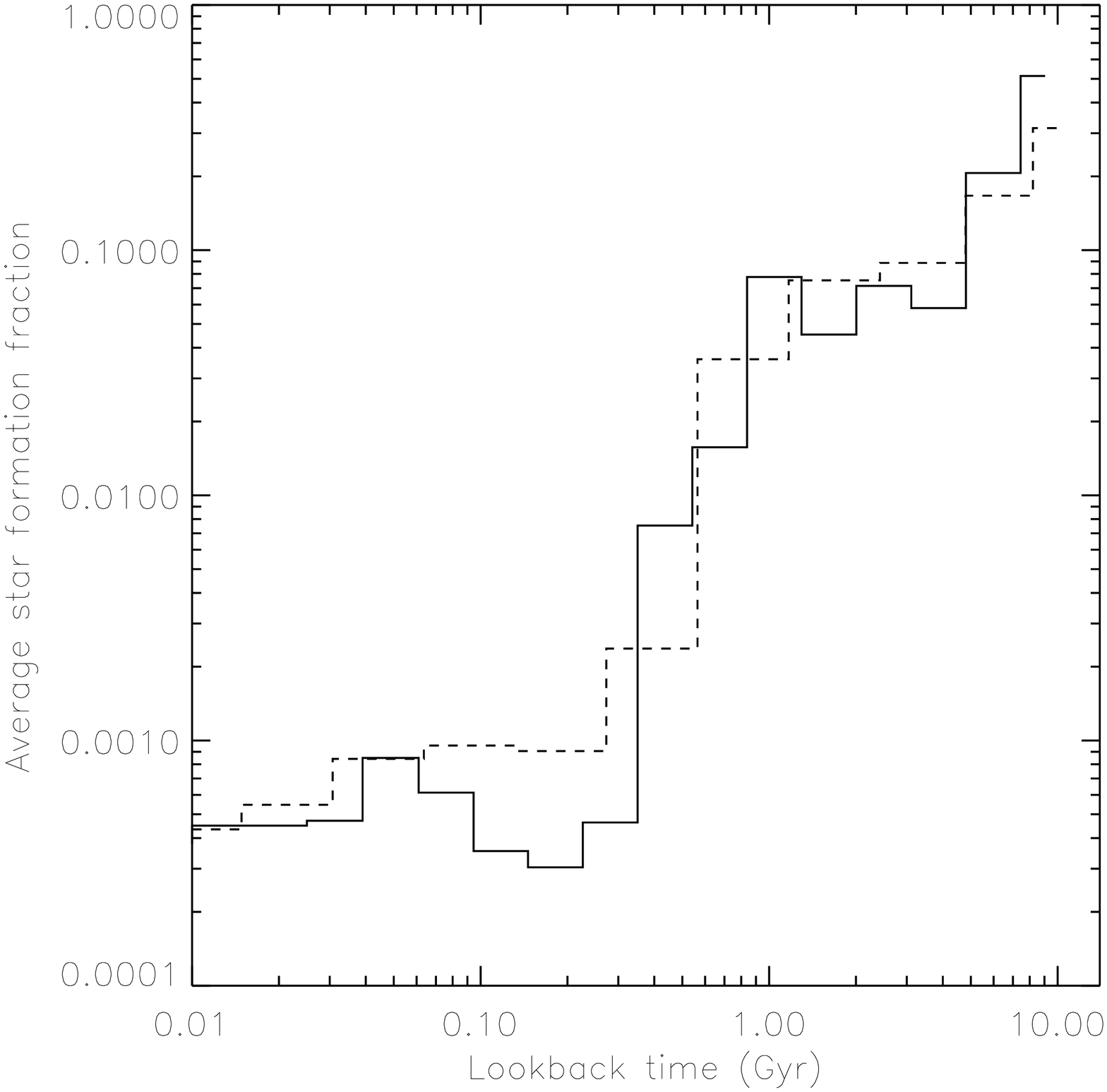}
\end{tabular}
\caption{The recovered average star formation history
  for the 821 galaxies as recovered by VESPA (solid line) and MOPED
  (dashed line). Both were initially normalised such that the sum over all bins
  is 1, and the MOPED line was then adjusted by 11/16 to account for
  the different number of bins used in each method, to facilitate
  direct comparison. 
}
\label{fig:averages}
\end{figure}
\par\noindent
Figure \ref{fig:averages} shows the recovered average star
formation history for the 821 galaxies using both methods. 
In the case of VESPA, solutions parametrized by low-resolution
bins had to be re-parametrized in high-resolution bins, so that a
common grid across all galaxies could be used. This was done using the weights given by
(\ref{eq:weights}). The lines show a remarkably good agreement between the two methods. \\
\par\noindent
Having recovered a star formation history for each galaxy, one can
then estimate the stellar mass of a galaxy. We calculated this
quantity for all galaxies using the solutions from both methods, and with
similar assumptions regarding cosmological parameters and fibre-size
corrections. Explicitly, we have done the following:
\begin{enumerate}
\item We converted from flux to luminosity assuming the set of
  cosmological parameters given by \cite{SpergelEtAl03MNRAS}.
\item We recovered the initial mass in each age bin using each method.
\item We calculated the remaining present-day mass for each population
  after recycling processes. This information is
  supplied by the synthetic stellar models, as a function of age and
  metallicity.
\item We summed this across all bins to calculate the total
  stellar present-day mass in the fibre aperture, $M$.
\item We corrected for the aperture size by scaling up the mass to
  $M_{stellar}$ using
  the petrosian and fibre magnitudes in the $z$-band, $M_p(z)$ and
  $M_f(z)$, with: $M_{stellar} = M \times 10^{0.4[M_p(z) - M_f(z)]}$
\end{enumerate}
\par
\begin{figure}
\vspace{0.6cm}
\begin{center}
\includegraphics[width=3.in]{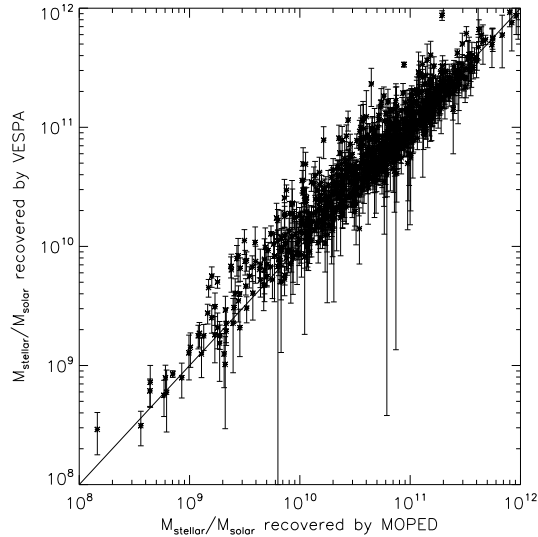}
\caption{Galaxy stellar mass (in units of solar masses) as recovered by VESPA and MOPED for
a sub-sample of 821 SDSS galaxies. The small percentage of galaxies with
significantly larger VESPA masses have large extinction.  The difference is
accounted for by the fact that MOPED and VESPA use different dust models.}
\label{fig:masses}
\end{center}
\end{figure}

\noindent
Figure \ref{fig:masses} shows the recovered galaxy masses as
recovered from MOPED and from VESPA. We see considerable agreement
between VESPA and MOPED. Over 75 per cent of galaxies have $ 0.5 \le
M_{VESPA}/M_{MOPED} \le 1.5$. There is a tail of around 10 per cent of
galaxies where VESPA recovers 2 to 4 times the mass recovered by
MOPED. The main reason for this difference is in the dust model used -
we find a correlation between dust extinction and the ratio of the two
mass estimates. This again reflects the fact that total stellar mass
estimates are highly sensitive to changes in the spectrum continuum
(see also section \ref{sec:duplicates}). \\ 
\par
\begin{figure}
\vspace{0.6cm}
\begin{center}
\includegraphics[width=3.in]{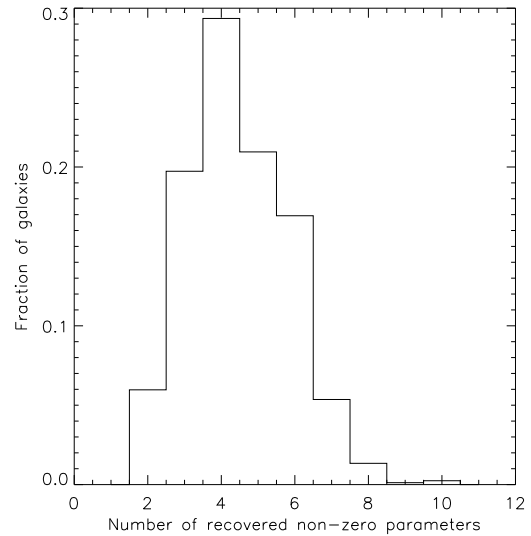}
\caption{Number of non-zero parameters in solutions recovered from 821
SDSS galaxies with VESPA. Please note that these correspond to the {\it total} number of non-zero
  components in the solution vector $\bf{c_\kappa}$ and not to the number
  of recovered stellar populations. For a information about the number
of recovered populations see Figure 14.}
\label{fig:Npars_SDSS}
\end{center}
\end{figure}

\noindent
Our sub-sample of 821 includes galaxies with a wide range of signal-to-noise ratios,
star formation histories and even wavelength range (mainly due to each
galaxy having different masks applied to it, according to the quality
of the spectroscopic data). Figure \ref{fig:Npars_SDSS} shows the
number of recovered non-zero parameters in the sample, using VESPA. As
an average, it falls below the synthetic examples studied in Section
3. This is not surprising, though, as each galaxy will have an unique
 and somewhat random combination of characteristics which will lead to
 a different number of parameters being recovered.  The total combination of
 these sets of characteristics would be impossible to investigate
 using the empirical method described in Section 3, and
 here lies the advantage of VESPA of dynamically adapting to each
 individual case. Also important to note is
 the fact that the wavelength coverage is normally not continuous in
 an SDSS galaxy, due to masked regions. This was not modelled in
 Section 3, and is likely to further reduce the number of recovered
 parameters in any given case. \\
\par
\begin{figure}
\vspace{0.6cm}
\begin{center}
\includegraphics[width=3.in]{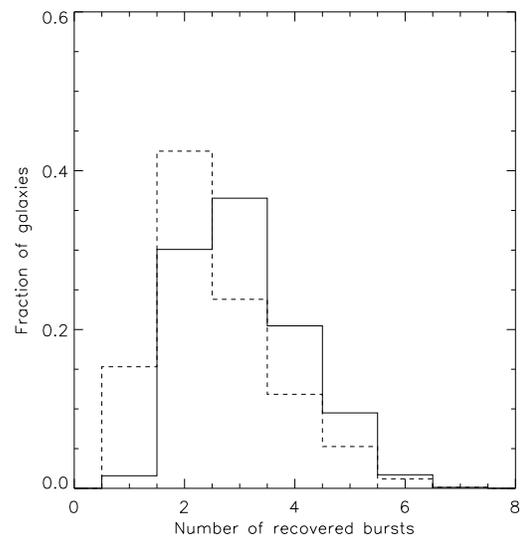}
\caption{The distribution of the total number of recovered stellar populations which
  contribute 5 per-cent or more to the total flux of the galaxy,
  as recovered from MOPED (dashed line) and VESPA (solid line).
}
\label{fig:Npops}
\end{center}
\end{figure}

\noindent
Perhaps more useful is to translate this number into a number of
recovered significant stellar populations for each galaxy. We define a
significant component as a stellar population which contributes 5
per cent or more to the total flux. Figure \ref{fig:Npops} shows the
distribution of the number of significant components for our
sub-sample of galaxies, as recovered by MOPED and VESPA. It is
interesting to note that both methods recover on average a similar
amount of components, even though MOPED has no explicit self-regularization
mechanism, as VESPA clearly does.

\section{Conclusions}
\label{sec:conclusions}        
We have developed a new method to recover star formation and
metallicity histories from integrated galactic spectra - VESPA.
Motivated by the current limitations of other methods which aim to do
the same, our goal was to develop an algorithm which is robust on a
galaxy-by-galaxy basis. VESPA works with a dynamic parametrization of
the star formation history, and is able
to adapt the number of parameters it attempts to recover from a given
galaxy according to its spectrum. In this paper we tested VESPA against a series of idealised synthetic
situations, and against SDSS data by comparing our results with those
obtained with the well-established code, MOPED. \\
\par\noindent
Using synthetic data we
found the quality and resolution of the recovered solutions varied with factors such
as type of star formation history, noise in the data and wavelength
coverage. In the vast majority of cases, and within the estimated errors and
bin-correlations, we observed a reliable reproduction of the input
parameters. As the signal-to-noise decreases, it becomes increasingly
difficult to recover robust solutions. Whereas our method cannot
guarantee a perfect solution, we have shown that the
self-regularization we imposed helped obtain a cleaner solution in
an overwhelming majority of the cases studied.\\
\par\noindent
On the real data analysis, we have studied possible effects from
systematics using duplicate observations of the same set of galaxies,
and have also compared VESPA's to MOPED's results obtained using the same
data sample. We found that in the majority of cases our results
are robust to possible systematics effects, but that in certain cases
and particularly when calculating stellar masses, VESPA might
underestimate the mass errors. However, we found no systematic
bias in any of our tests. We have also shown that VESPA's results are in
good agreement with those of MOPED for the same sample of
galaxies. VESPA and MOPED are two fundamentally different approaches
to the same problem, and we found good agreement both in a global
sense by looking at the average star formation history of the sample,
and in an individual basis by looking at the recovered stellar masses
of each galaxy. VESPA typically recovered between 2 to 5 stellar
populations from the SDSS sample. \\
\par\noindent
VESPA's ability to adapt dynamically to each galaxy and to extract
only as much information as the data warrant is a completely new way
to tackle the problem of extracting information from galactic
spectra. Our claim is that, for the most part, VESPA's results are
robust for any given galaxy, but our claim comes with two words of
caution. The first one concerns very noisy galaxies - in extreme
cases (SNR$\approx$10 or less, at a resolution of $3 \AA$), it becomes very difficult to extract
any meaningful information from the data. This uncertainty is evident
in the large error bars and bin-correlations, and the solutions can be
essentially unconstrained even at low-resolutions. We are therefore
limited when it comes to analysing individual high-noise galaxies,
which is the case of many SDSS objects. Our
second word of caution concerns the stellar models used to
analyse real galaxies - any method can only do as well as the models it
bases itself upon. We are limited in our knowledge and ability to
reproduce realistic synthetic models of stellar populations, and this
is inevitably reflected in the solutions we obtain by using them. On
the plus side, VESPA works with any set of synthetic models and can take
advantage of improved versions as they are developed. \\
\par\noindent
VESPA is fast enough to use on large spectroscopic samples (a
typical SDSS galaxy takes 1 minute on an average workstation), and we
are in the process of analysing SDSS's Data Release 5 (DR5),
which consists of roughly half a million galaxies. Our first aim is to
publish and exploit a catalogue of robust star formation histories,
which we hope will be a valuable resource to help constrain models of
galaxy formation and evolution.  

\section{Acknowledgments}
We are grateful to the anonymous referee for a very thoughtful report which led to material improvements in the paper.\\
\par\noindent
RT is funded by the Funda\c{c}\~{a}o para a Ci\^{e}ncia e a
Tecnologia under the reference PRAXIS SFRH/BD/16973/04. RJ's research is 
supported by the NSF through grant PIRE-0507768 and AST-0408698 to the Atacama
Cosmology Telescope.\\
\par\noindent
    Funding for the SDSS and SDSS-II has been provided by the Alfred
    P. Sloan Foundation, the Participating Institutions, the National
    Science Foundation, the U.S. Department of Energy, the National
    Aeronautics and Space Administration, the Japanese Monbukagakusho,
    the Max Planck Society, and the Higher Education Funding Council
    for England. The SDSS Web Site is http://www.sdss.org/. The SDSS is managed by the Astrophysical Research Consortium for the Participating Institutions. The Participating Institutions are the American Museum of Natural History, Astrophysical Institute Potsdam, University of Basel, University of Cambridge, Case Western Reserve University, University of Chicago, Drexel University, Fermilab, the Institute for Advanced Study, the Japan Participation Group, Johns Hopkins University, the Joint Institute for Nuclear Astrophysics, the Kavli Institute for Particle Astrophysics and Cosmology, the Korean Scientist Group, the Chinese Academy of Sciences (LAMOST), Los Alamos National Laboratory, the Max-Planck-Institute for Astronomy (MPIA), the Max-Planck-Institute for Astrophysics (MPA), New Mexico State University, Ohio State University, University of Pittsburgh, University of Portsmouth, Princeton University, the United States Naval Observatory, and the University of Washington.


\appendix
\end{document}